# Spin Networks and Quantum Gravity

Carlo Rovelli* and Lee Smolin†

* *Department of Physics, University of Pittsburgh, Pittsburgh, Pa15260*
† *Center for Gravitational Physics and Geometry, Department of Physics*
*Pennsylvania State University, University Park, Pa16802-6360*
*and*
*School of Natural Science, Institute for Advanced Study,*
*Princeton, NJ 08540*

April 24, 1995

**Abstract**

We introduce a new basis on the state space of non-perturbative quantum gravity. The states of this basis are linearly independent, are well defined in both the loop representation and the connection representation, and are labeled by a generalization of Penrose's spin networks. The new basis fully reduces the spinor identities (SU(2) Mandelstam identities) and simplifies calculations in non-perturbative quantum gravity. In particular, it allows a simple expression for the exact solutions of the Hamiltonian constraint (Wheeler-DeWitt equation) that have been discovered in the loop representation. Since the states in this basis diagonalize operators that represent the three geometry of space, such as the area and volumes of arbitrary surfaces and regions, these states provide a discrete picture of quantum geometry at the Planck scale.

* rovelli@vms.cis.pitt.edu, † smolin@phys.psu.edu

## 1 Introduction

The loop representation [1, 2] is a formulation of quantum field theory suitable when the degrees of freedom of the theory are given by a gauge field, or a connection. This formulation has been used in the context of continuum and lattice gauge theory [3], and it has found a particularly effective application in quantum gravity [2, 4], because it allows a description of the diffeomorphism invariant quantum states in terms of knot theory [2, 5], and, at the same time, because it partially diagonalizes the quantum dynamics of the theory, leading to the discovery of solutions of the dynamical constraints [2, 6]. Recent results in quantum gravity based on the loop representation include the construction of a finite physical Hamiltonian operator for pure gravity [7] and fermions [8], the computation of the physical spectra of area [9] and volume [10], and the developement of a perturbation scheme that may allow transition amplitudes to be explicitely computed [7, 11, 12]. A mathematically rigorous formulation of quantum field theories whose configuration space is a space of connections, inspired by the loop representation, has been recently developed [13, 14] and the kinematics of the theory is now on a level of rigor comparable to that of constructive quantum field theory [15]. This approach has also produced interesting mathematical spin-off's such as the construction of diffeomorphism invariant generalised measures on spaces of connections [14] and could be relevant for a constructive field theory approach to non-abelian Yang-Mills theories.

Applications of the loop representation, however, have been burdened by complications arising from two technical nuisances. The first is given by the Mandelstam identities, because of which the loop states are not independent and form an overcomplete basis. The second is the presence of a certain sign factor in the definition of the fundamental loop operators $T^n$ for $n > 1$. This sign depends on the global connectivity of the loops on which the operator acts and obstructs a simple local graphical description of the operator's action. In this work, we describe an elegant way to overcome both of these complications. This comes from using a particular basis, which we denote as spin network basis, since it is related to the spin networks of Penrose [16]. The spin network basis has the following properties.

- i. It solves the Mandelstam identities.
- ii. It allows a simple and entirely local graphical calculus for the $T^n$ operators.

- iii. It diagonalises the area and volume operators.

The spin network basis states, being eigenstates of operators that correspond to measurement of the physical geometry, provide a physical picture of the three dimensional quantum geometry of space at the Planck-scale level.

The main idea behind this construction, long advocated by R. Loll [17], is to identify a basis of independent loop states in which the Mandelstam identities are completely reduced. We achieve such a result by exploiting the fact that all irreducible representations of $SU(2)$ are built by symmetrized powers of the fundamental representation. We will show that in the loop representation this translates into the fact that we can suitably antisymmetrize all loops overlapping each other, without loosing generality. More precisely, the (suitably) antisymmetrized loop states span, but do not over span, the kinematical state space of quantum gravity.

The independent basis states constructed in this way turn out to be labelled by Penrose's spin networks [16], and by a direct generalization of these. A spin network is a graph whose links are "colored" by integers satisfying simple relations at the intersections. Roger Penrose introduced spin networks in a context unrelated to the present one; remarkably, however, his aim was to explore a quantum mechanical description of the geometry of space, which is the same ambition that underlies the loop representation construction.

The idea of using a spin network basis has appeared in other contexts in which holonomy of a connection plays a role, including lattice gauge theory, [18, 19] and topological quantum field theory [20, 21, 22, 23, 24, 25]. The use of this basis in quantum gravity has been suggested previously [26], but its precise implementation had to await resolution of the sign difficulties mentioned above. Here, these difficulties are solved by altering a sign in the relation between the graphical notation of a loop and the corresponding quantum state. This modified graphical notation for the loop states allows us to reduce the loop states to the independent ones by simply antisymmetrizing overlapping loops.

The spin-network construction has already suggested several directions of investigation, which are being pursued at the present time. The fact that it diagonalizes the operator that measures the volume of a spatial slice [10] gives us a physical picture of a discrete quantum geometry and also makes the spin network basis useful for perturbation expansions of the dynamics of general relativity, as described in [7, 11, 12]. It has also played a role in the mathematically rigorous investigations of refs. [15, 27]. Another intriguing suggestion is the possibility of considering $q$ deformed spin-networks, on which we will comment in the conclusion.

The details of the application of the spin network basis to the diagonalization of the volume and area operators have been described in an earlier paper [10]. The primary aim of this paper is to give an introduction to the spin network basis and to its use in nonperturbative quantum gravity. We emphasize the details of its construction, at a level of detail and rigor that we hope will be useful for practical calculations in quantum gravity. No claims are made of mathematical rigor; for that we point the reader to the recent works by Baez [28] and Thiemann [27], where the spin network basis is put in a rigorous mathematical context. Finally, we note that in this paper we work with $SL(2,C)$ (or $SU(2)$) spinors, which are relevant for the application to quantum gravity, but a spin network basis such as the one we describe exists for all compact gauge groups [28].

This paper is organized as follows. In the next section we briefly explain the two problems that motivate the use of the spin network basis. This leads to section 3, in which we provide the definition of spin network states in the loop representation. In section 4, we describe the spin network states as they appear in the connection representation [29]. The proof that the spin network states do form a basis of independent states may then be given in section 5. Following this, in section 6, we review the general structure of the transformation theory (in the sense of Dirac) between the loop representation and the connection representation. The use of the spin network basis considerably simplifies the transformation theory, as we show here. Similarly, old results on the existence of solutions to the hamiltonian constraint and exact physical states of quantum gravity may be expressed in a simpler way in terms of the spin network basis. Its use makes it unnecessary to explicitly compute the extensions of characteristic states of nonintersecting knots to intersecting loops, as described in [2, 26][1].

Finally, an important side result of the analysis above is that it indicates how to modify the graphical calculus in loop space in order to get rid of the annoying non-locality due to the dependence on global rooting. The new notation that allows a fully local calculus is defined in section 7. The paper closes with a brief summary of the results in section 8, and with a short

[1] These characteristic states were previously defined to be equal to one on the knot class of a non-intersecting loop, zero on all other non-intersecting loops, with an extension to the classes of intersecting loops defined by solving the Mandelstam identities [2, 26]. Now they may be succinctly described as being equal to one on one element of the spin network basis and zero on all the others.

appendix in which we discuss the details of the construction of higher than trivalent vertices.

## 2 Definition of the problem

The loop representation is defined by the choice of a basis of bra states $\langle\alpha|$ on the state space of the quantum field theory. These states are labelled by loops $\alpha$. By a loop, we mean here a set of a finite number of single loops; by a single loop, we mean a piecewise smooth map from the circle $S^1$ into the space manifold. The loop basis is characterised (defined) by the action on this basis of a complete algebra of observables [2]. A quantum states $|\psi\rangle$ is represented in this basis by the loop space function

$$\psi(\alpha) = \langle\alpha|\psi\rangle. \tag{1}$$

For detailed introductions and notation we refer to [26, 29, 30]. As shown in [1, 2], the functions $\psi(\alpha)$ that represent states of the system must satisfy a set of linear relations, which we denote as the Mandelstam relations. These code, among other things, the structure group of the Ashtekar's connection [31] $A$ of the classical theory. Let $U_\alpha(A) = exp\int_\alpha A$ be the parallel propagator matrix, or holonomy, of the connection $A$ along the curve $\alpha$, and let $T[A,\alpha] = TrU_\alpha(A)$ be its trace. Then the Mandelstam relations are defined, for the present purposes, as follows[13]. For every set of loops $\alpha_1, ...\alpha_N$ and complex numbers $c_1, ..., c_N$ such that

$$\sum_k c_k\, T[A,\alpha_k] = 0 \tag{2}$$

holds for all (smooth) connections $A$, the states $\psi(\alpha)$ must satisfy

$$\sum_k c_k\, \psi(\alpha_k) = 0. \tag{3}$$

It follows that the states of the bra basis $\langle\alpha|$ are not independent (as linear functionals on the ket state space), but satisfy the identities

$$\sum_k c_k \langle\alpha_k| = 0. \tag{4}$$

The basis $\langle\alpha|$ is therefore overcomplete. Let us from now on concentrate on the $SL(2,C)$ case. There are two cases of the relation (4) that are particularly interesting. The first one yields the spinor identitity, or proper $SL(2,C)$ Mandelstam identy: Given any two $SL(2,C)$ matrices $A$ and $B$, the following holds between the traces we can construct in terms of them.

$$Tr(A)Tr(B) - Tr(AB) - Tr(AB^{-1}) = 0. \tag{5}$$

Let $\alpha$ and $\beta$ be two loops that intersect in a point $p$. Let $\alpha \cdot \beta$ and $\alpha \cdot \beta^{-1}$ be the two loops that are obtained by starting at $p$, going around $\alpha$ and then around either $\beta$ or $\beta^{-1}$, so that $U_{\alpha\cdot\beta} = U_\alpha U_\beta$ and $U_{\alpha\cdot\beta^{-1}} = U_\alpha (U_\beta)^{-1}$ Then (5) implies that for every $A$,

$$T[A, \alpha \cup \beta] - T[A, \alpha \cdot \beta] - T[A, \alpha \cdot \beta^{-1}] = 0. \tag{6}$$

Therefore, we have

$$\langle \alpha \cup \beta| - \langle \alpha \cdot \beta| - \langle \alpha \cdot \beta^{-1}| = 0. \tag{7}$$

The second example which is easily seen deriving from (4) is the retracing identity

$$\langle \alpha \cdot \gamma \cdot \gamma^{-1}| = \langle \alpha|, \tag{8}$$

where $\gamma$ is an open segment with an end point on the loop $\alpha$ (a "tail"). In earlier work [2] it has been assumed that all identities (4) can be derived from the two relations (7) and (8). We are not aware of any complete proof, or of a counterexample of this conjecture.

The redundancy introduced in the loop representation by the Mandelstam identities is cumbersome. It is not the spinor identity by itself, nor the retracing identity by itself that create much complications, since the first could be solved by simply choosing a list of independent intersections, and the second by discarding all loops with "tails" from the theory. It is the *combination* of the two relations which makes it difficult to isolate a set of independent loop functionals. To see this, consider two loops $\alpha$ and $\beta$ that do not intersect. At first sight, one would say that these are not affected by the spinor relation, but they are. To see how, consider an open segment $\gamma$ with one end on $\alpha$ and one end on $\beta$. Combining (7) and (8) we have

$$\langle \alpha \cup \beta| - \langle \alpha \cdot \gamma \cdot \beta \cdot \gamma^{-1}| - \langle \alpha \cdot \gamma \cdot \beta^{-1} \cdot \gamma -1| = 0. \tag{9}$$

So that even non-intersecting multiple loop state enters the Mandelstam identities. Equation (9) can be represented graphically as

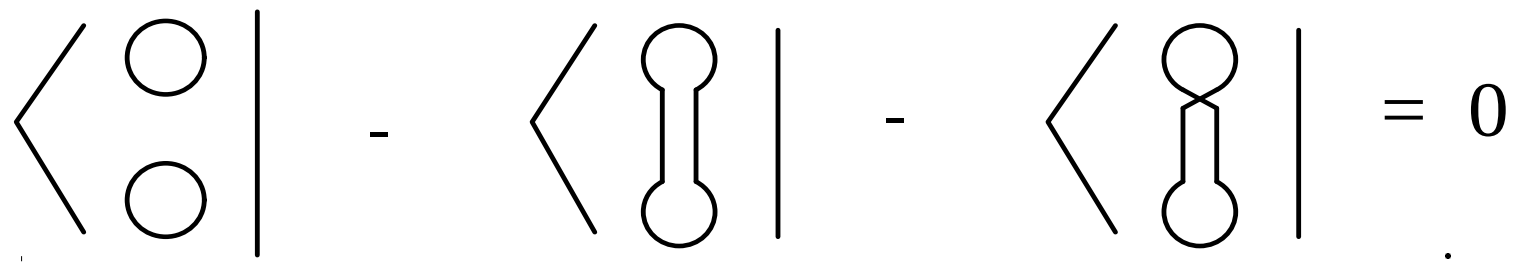

(10)

The problem that we consider in this paper is to find a basis of states $\langle s|$ that are fully independent, so that no linear combination of them can be set to zero using the identities (4). Such a basis will be defined in section 3, and the proof of indipendence given in section 5. The rest of this section describes the motivations underlying the definitions in section 3.

## 2.1 The sign difficulty

There is a natural strategy for getting rid of the redundancy expressed in eq. (10), which we are now going to describe. This strategy, however, is obstructed by a sign difficulty, which previously prevented its complete implementation. The natural strategy is to get rid of the degeneracy by antisymmetrizing all lines running parallel to each other. For instance, out the tree loop states involved in relation (9), we may pick the two independent states

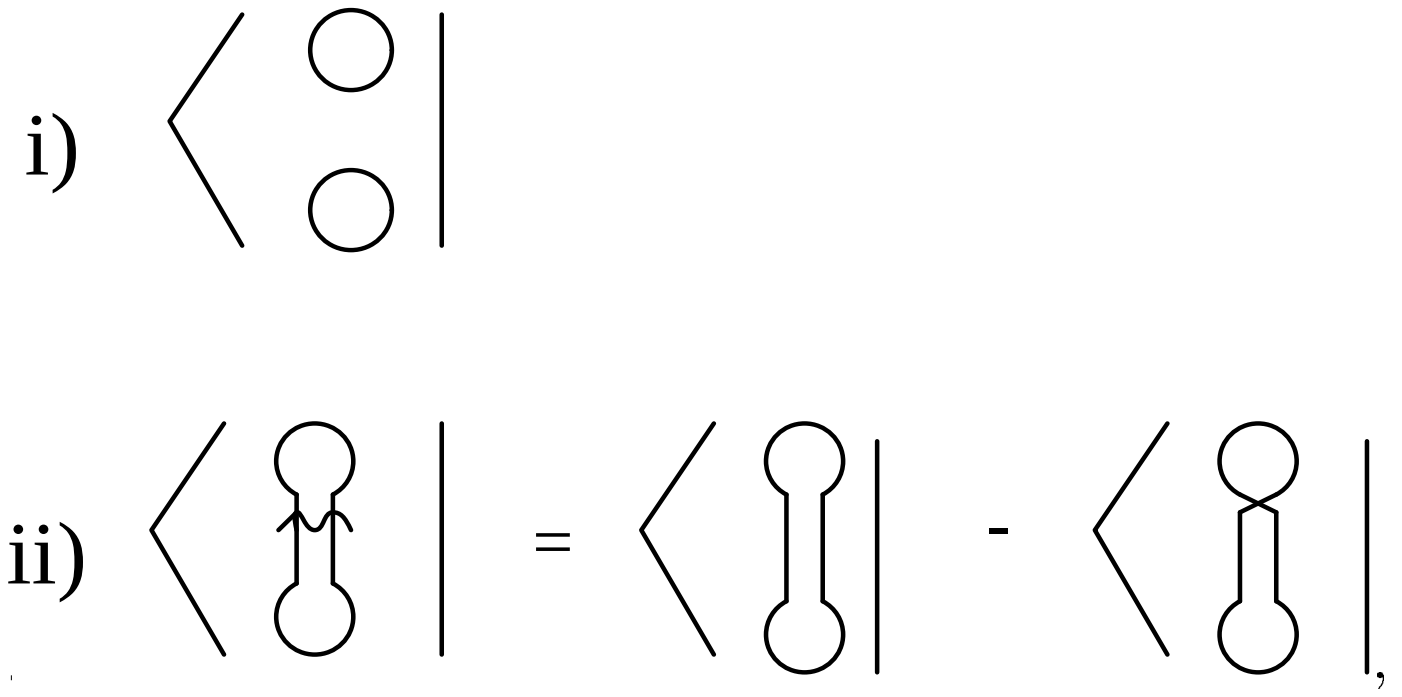

(11)

where we have indicated antisymmetrization by a wiggly line. Since the symmetric combination $\langle\alpha\cdot\gamma\cdot\beta\cdot\gamma^{-1}| + \langle\alpha\cdot\gamma\cdot\beta^{-1}\cdot\gamma{-}1|$ is equal to $\langle\alpha\cup\beta|$ by equation (9), the two states above exhaust all possible independent loop states that can be constructed out of three original states. This procedure should be combined with some suitable restriction to the independent intersections.

Let us introduce some terminology. We denote a set of loop segments that fully overlap as a "rope", and we call the number of loops that form it, without regard to orientation, the "order of the rope". Thus $\gamma$ and $\gamma^{-1}$ form a rope of order two in the second and third states in ii) above. Given an intersection point $p$ of the loop (a point on the support of the loop where this support fails to be a submanifold of $\Sigma$), we denote the number of ropes that emerge from $p$ as the order of the intersection; and we say that a loop is $n-$valent if it has intersections of order at most $n$. To begin with, we shall only consider trivalent loops. For instance, in the example above the intersection between $\alpha$ and $\gamma$ in the loop $\alpha{\cdot}\gamma{\cdot}\beta{\cdot}\gamma^{-1}$ is trivalent because $\gamma$ and $\gamma^{-1}$ form a single set of overlapping loop segments (a single rope) emerging from the intersection. We will deal with non-trivalent intersections in the Appendix.

We may hope to reduce the degeneracy by replacing every overlapping segment with a suitable antisymmetrized combination, plus "tails" that can be got rid of by means of the retracing identity. In the example considered above, for instance, we can reduce the state $\langle \alpha \cdot \gamma \cdot \beta \cdot \gamma^{-1}|$ to a linear combination of the two states defined in (11) as follows

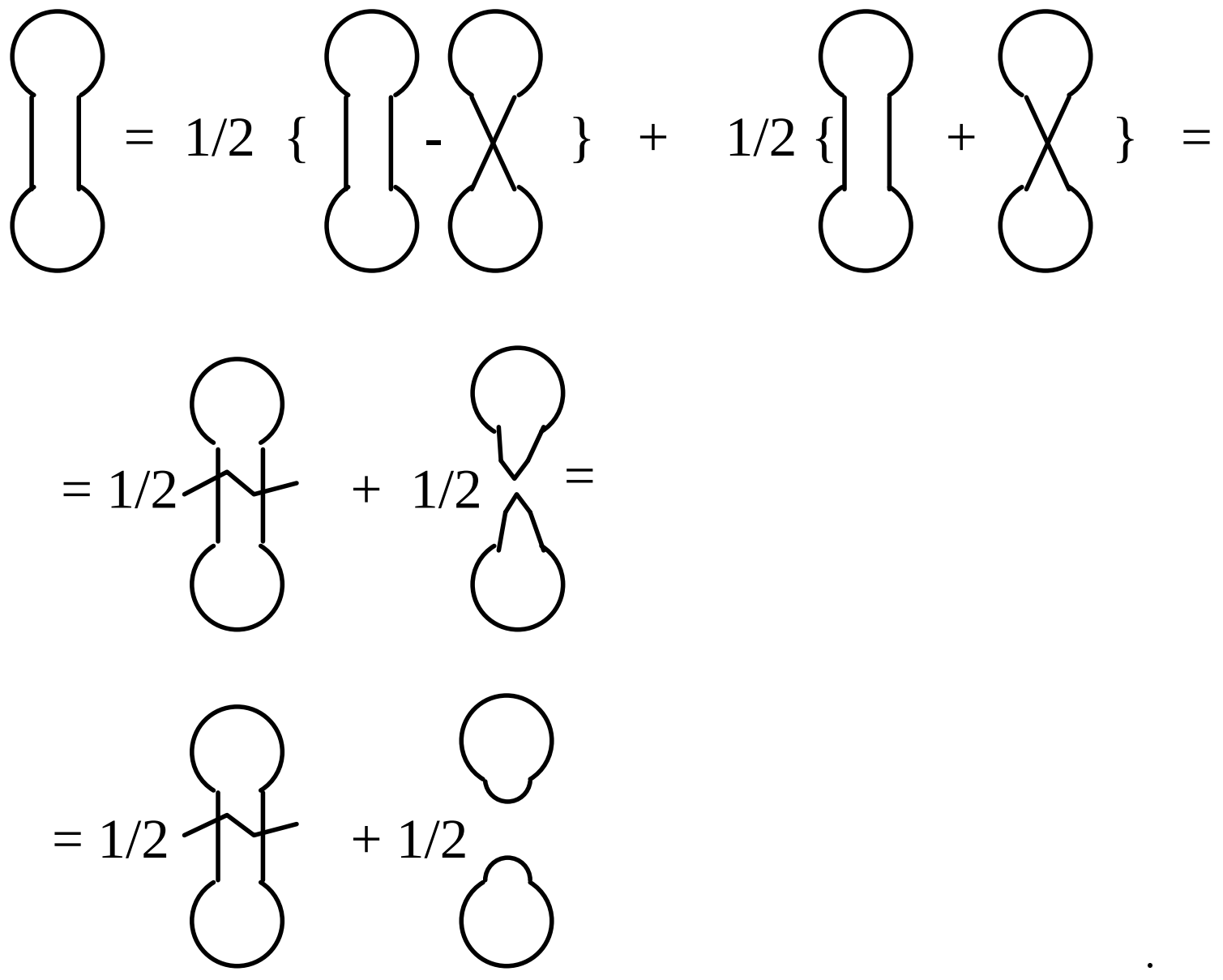

(12)

So we may hope that any time we have two parallel lines, we could use the spinor identity as follows

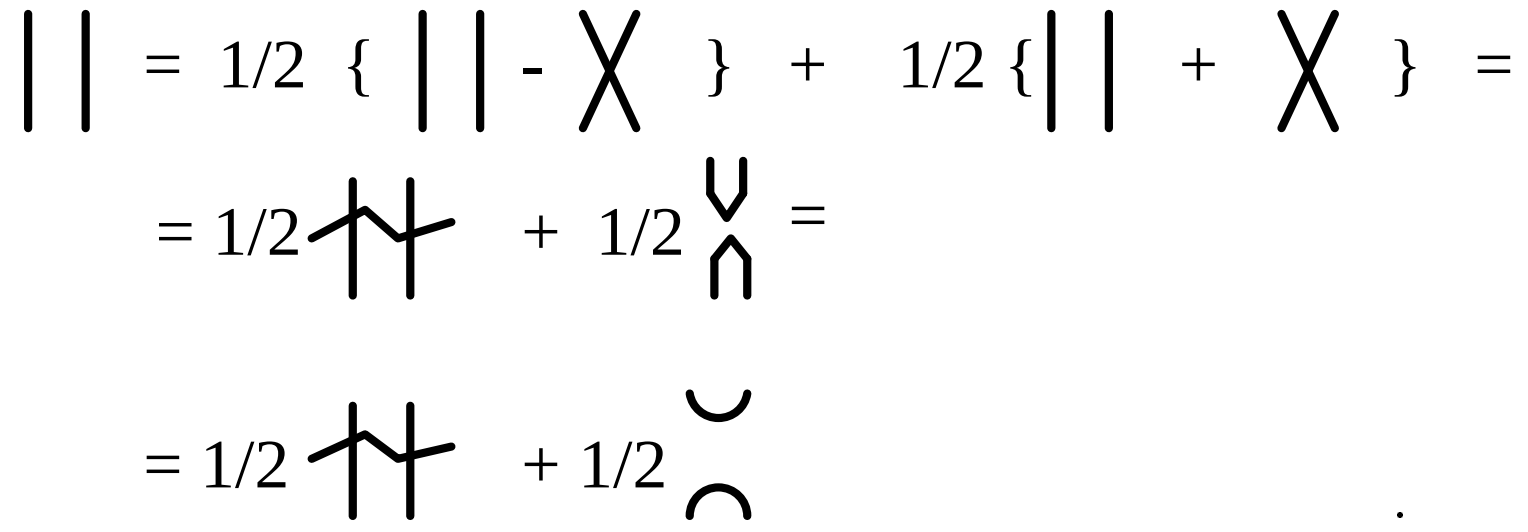

(13)

Unfortunately, this does not work. To understand why, consider a loop $\alpha$ and an open segment $\gamma$ that starts and ends in two different points of $\alpha$. Denote by $\alpha_1$ and $\alpha_2$ the two segments in which the two intersections with $\gamma$ partition $\alpha$. Then we have, due to the spinor and retracing identities

$$\langle\alpha| - \langle\alpha_1\cdot\gamma^{-1}\cup\alpha_1\cdot\gamma| + \langle\alpha_1\cdot\gamma\cdot\alpha_1\cdot\gamma| = 0; \tag{14}$$

namely

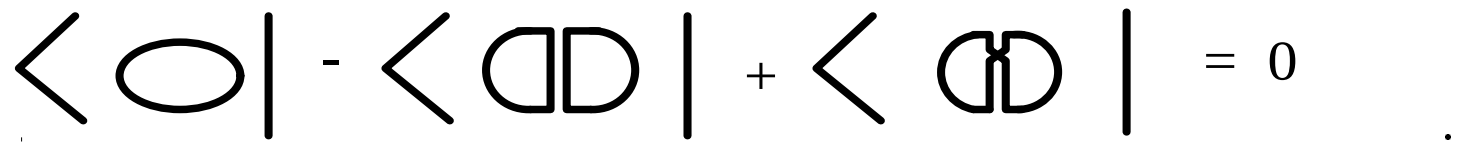

(15)

If we want to pick two independent linear combinations, we have to chose the *symmetric* combination $\langle\alpha| + \langle\alpha_1\cdot\gamma^{-1}\cup\alpha_1\cdot\gamma|$, and not the *antisymmetric* one as before. Namely, we have to choose

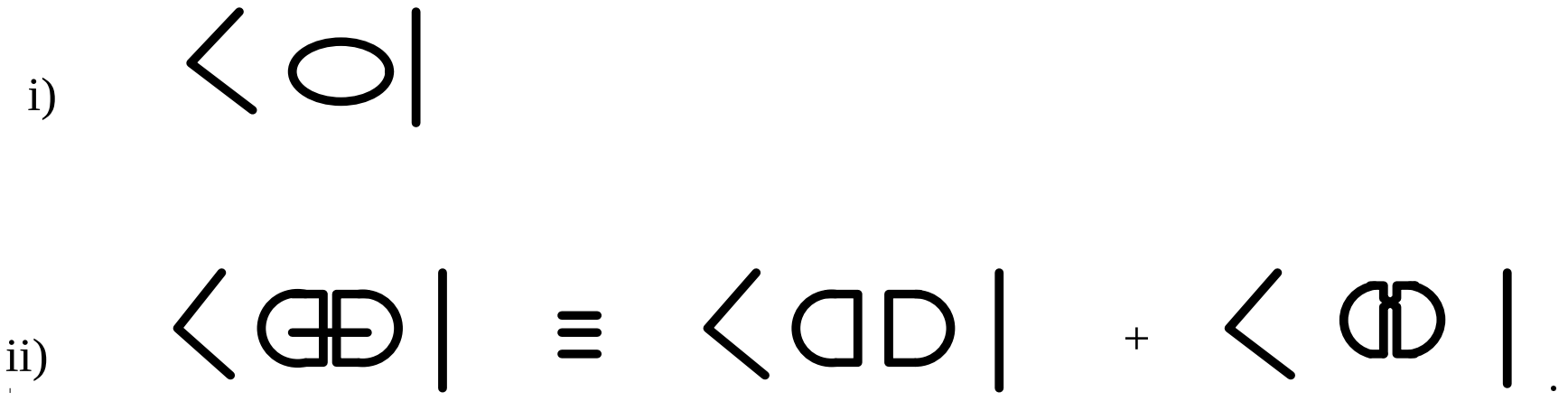

(16)

Thus, to pick the independent combination of loop states, we have to antisymmetrize the rope in one case, but we have to symmetrize it in the other case. In general, the choice between symmetrization and antisymmetrization can only be worked out by writing out explicitly the full pattern of rootings in the multiple loop. In other words, equation (13) is in general wrong if taken as a calculation rule that can be used in dealing with any loop state. More precisely: at every intersection, the spinor identity provides a linear relation between the three multiple loops obtained by replacing the intersection with the three possible rootings through the loop

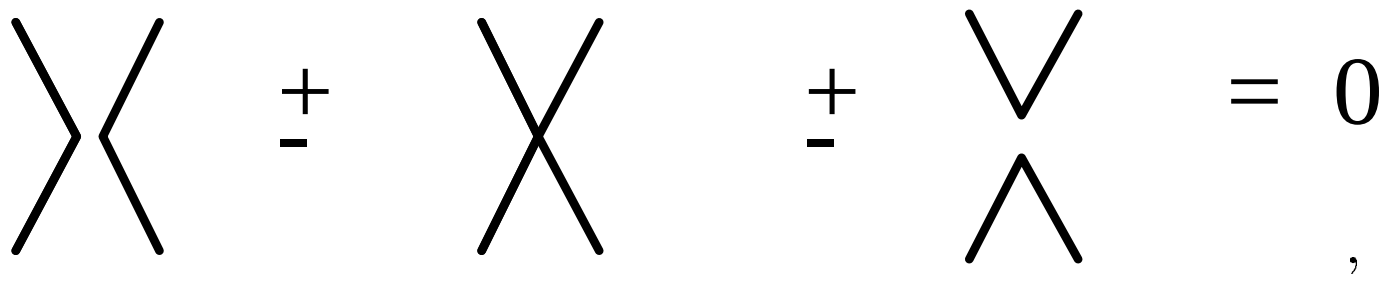

(17)

but the sign in front of each term depends on the global rooting of the loops.

There is a simple way out of this difficulty, which does allow us to get rid of the spinor identities among trivalent loops simply by antisymmetrization. In order to determine the correct signs of the various terms in eq. (17), we have to take the global rooting into account. There are only three possibilities:

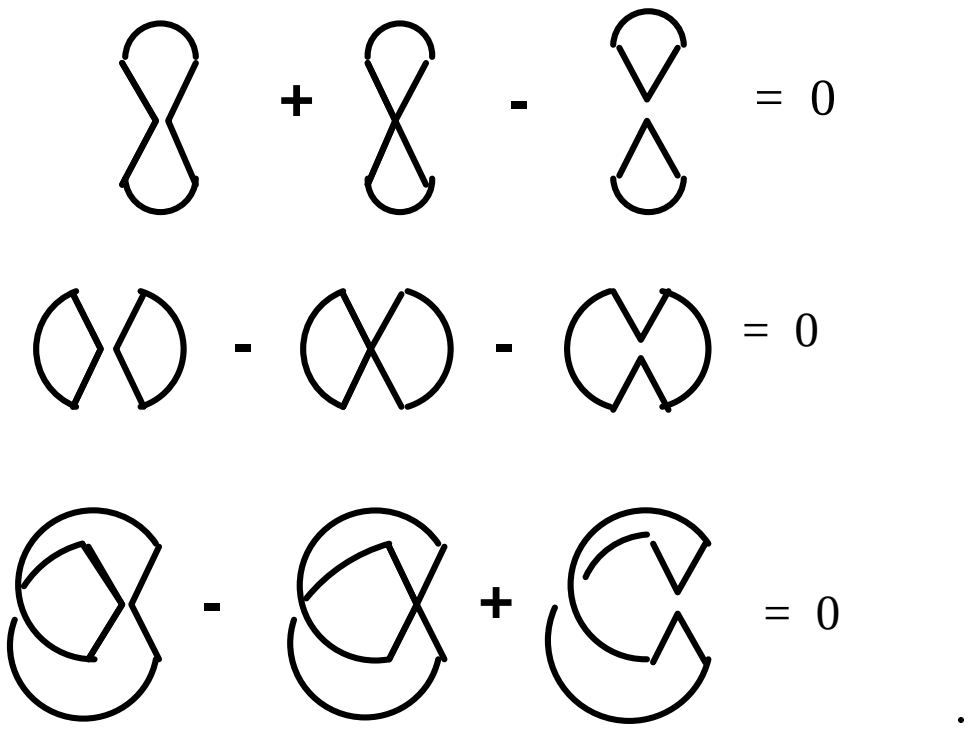

(18)

The signs derive immediately from eq.(7). Now, the way out of the sign difficulty is provided by the following observation. If we multiply each term $\gamma$ of the linear combinations in eq. (18) by the sign-factor

$$\epsilon = (-1)^{n(\gamma)} \qquad (19)$$

where $n(\gamma)$ is the number of single loops in the term $\gamma$, we obtain the following relations.

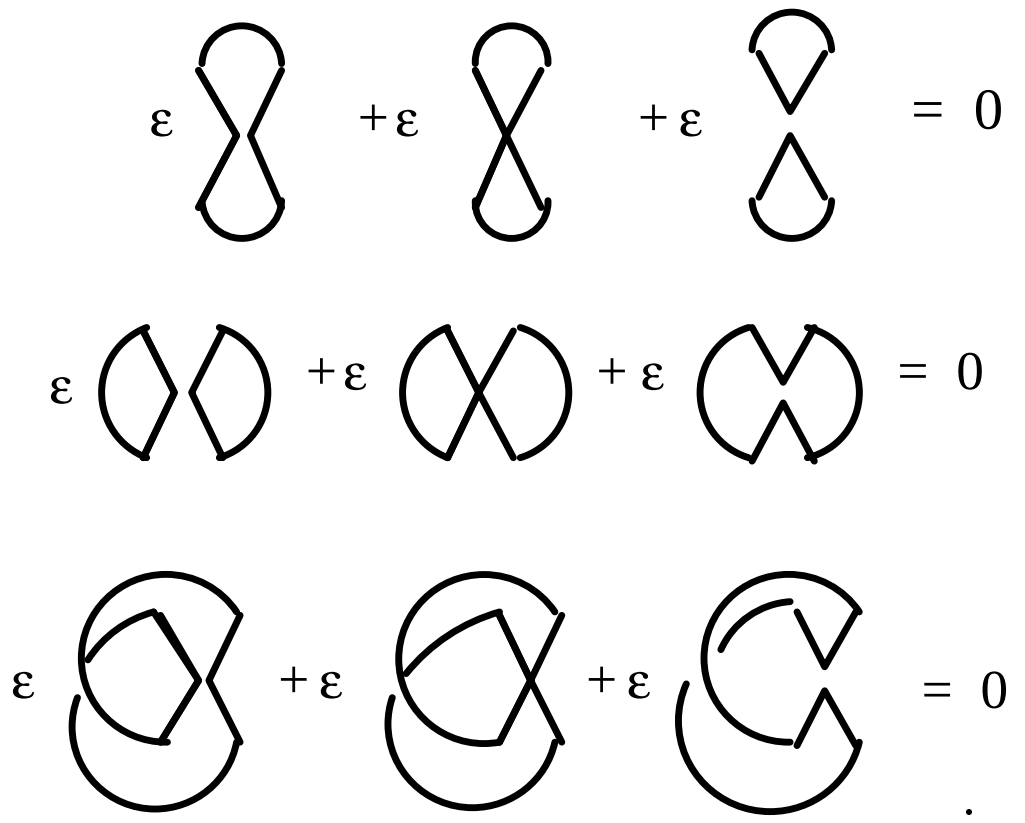

(20)

Thus, if we multiply all terms by $(-1)^{n(\gamma)}$, we can use the algebra of eq. (13), and therefore reduce every overlapping loop to fully antisymmetrized terms plus terms where two overlapping disappear by means of the retracing identity. In other words, the indipendent states must be constructed by fully antisimmetrizing the segments along the ropes and multiplying the resulting terms by $(-1)^{n(\gamma)}$.

Let us study the set of states determined in this way. It is easy to convince oneself that if the three ropes adjacent to a trivalent node are completely symmetrized, then the rootings of the single loop-segments through the intersection are uniquely determined. It follows that the (trivalent) states that we have obtained by antisymmetrizing the ropes are fully determined solely by their support, the order of each rope and an overall sign. Equivalently, they are determined by a trivalent graph (the support), with integers assigned to each link (the order of the rope), plus an orientation of the graph. Furthermore, the orders of the three ropes adjacent to a given node are constrained to satisfy some relations among themselves. First, we can assume that no loop through the node can go back to the rope it comes from (otherwise we can retrace it away). Thus there are three sets of loops that run through a trivalent intersection: the ones rooted from the first to the second rope (let's say we have $a$ of them), the ones rooted from the second to the third rope ($b$ of them), and the ones rooted from the third to the first rope ($c$ of them). It follows that the order of the three ropes are, respectively:

$$p = c + a, \quad q = a + b, \quad r = b + c. \tag{21}$$

The three numbers $a, b$ and $c$ are arbitrary positive integers, but not so the orders $p, q$ and $r$ of the adjacent ropes. It follows immediately from (21) that they satisfy two relations:

- i. Their sum is even,
- ii. None is larger than the sum of the other two;

and that these two conditions on $p, q$ and $r$ are sufficient for the existence of $a, b$ and $c$. We conclude that our states are labelled by oriented trivalent graphs, with integers $p_l$ associated to each links $l$, such that at every node the relations i. and ii. are satisfied. By definitions, these are Penrose's spin networks [16]. Thus, a linear combination of trivalent loops with the same support, in which every rope is fully antisymmetrized is uniquely determined by an an imbedded, oriented, trivalent spin network. We shall denote these fully antisymmetrized states as spin-network states. From the discussion we have just had we can see that they comprise an independent basis.

Using the above discussion as motivation, in the next section we provide a complete definition of spin networks and spin network quantum states.

## 3 Spin-network states in the loop representation

In this section we define the spin network states and their corresponding diffeomorphism invariant knot states. As defined by Penrose, a spin network is a trivalent graph, $\Gamma$, in which the links $l$ are labled by positive integers $p_l$, denoted "the color of the link", such that the sum of the colors of three links adjacent to a node is even and none of them is larger than the sum of the other two. To each spin network we may associate an orientation, $+1$ or $-1$, determined by assigning a cyclic ordering to the three lines emerging from each node. In particular, an orientation is determined by a planar representation of the graph (by the clockwise ordering of the lines), and gets reversed by redrawing one of the intersection with two lines emerging in inverted order. Here we consider imbedded, oriented, spin networks, and we denote such objects by the capitalized latin letters $S, T, R...$ . An imbedded spin network is a spin network plus an immersion of its graph in the three-dimensional manifold $\Sigma$. Later, in discussing the solution of the diffeomorphism constraint, we will consider equivalence classes of these spin networks under diffeomorphisms; these will be called "s-knots" and indicated by lower case latin letters $s, t, r, ...$ .

Given a trivalent imbedded oriented spin network (from now on, just spin network) $S$, we can construct a quantum state of the loop representation as follows. First we replace every link $l$ of the spin network by a rope of degree $p$, where $p$ is the color of the link $l$. Then, at every intersection we join the segments that form the rope pairwise, in such a way that each segment is joined with one of the segments of a different rope. As illustrated in the previous section, the constraints on the coloring turn out to be precisely the necessary and sufficient conditions for the matching to be possible. The matching produces a (multiple) loop, which we denote as $\gamma_1^S$. Then, we consider the $M = \prod_l p_l!$ loops $\gamma_m^S, m = 1, ..., M$ that can be obtained from $\gamma_1^S$ by permutations of the loops along each rope. (Each rope of color $p_l$ produces $p_l!$ terms.) We assign to each of these loops a sign factor $(-1)^{c(m)}$ which is positive/negative for even/odd permutations of the loops in $\gamma_1^S$. (Equivalently, one can identify $c(m)$ with the number of crossings along ropes, in a planar representation of the loops.) Finally we define the state

$$\langle S| \equiv \sum_m (-1)^{c(m)} (-1)^{n(m)} \langle \gamma_m^S| \tag{22}$$

(here and in he following we write $n(\gamma_m^S)$ as $n(m)$ for short; we recall that $n(\gamma)$ is the number of single loops forming the multiple loop $\gamma$). We denote the state $\langle S|$ defined by equation (22) as spin network state, or the quantum state associated with the spin network $S$.

Notice that, up to the overall sign, the linear combinaton that defines $\langle S|$ is independent on the particular rooting through the intersections chosen in constructing $\gamma_1^S$, because every other rooting is produced by the permutations. The overall sign is fixed by the orientation of the spin network. For concreteness, let us assign an orientation to the spin network by projecting it on a plane, and assign $(-1)^{c(m)} = 1$ to the (unique) loop $\gamma_1^S$ among the $\gamma_m^S$ that can be drawn without crossings the segments ($c(1) = 0$) along the ropes and in the nodes. We will show in section 4 that the states $\langle S|$ we have defined form a basis of independent states for the trivalent quantum states.

We represent spin network states simply by drawing their graphs and labeling the edges with the corresponding colors, and, if necessary, with the name of the loop or segment they correspond to. As an example, and in order to illustrate how the signs are taken into account by the above definitions, consider the spin network

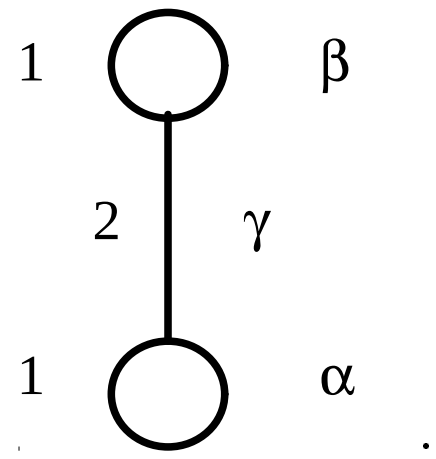

(23)

This is expanded in loops as

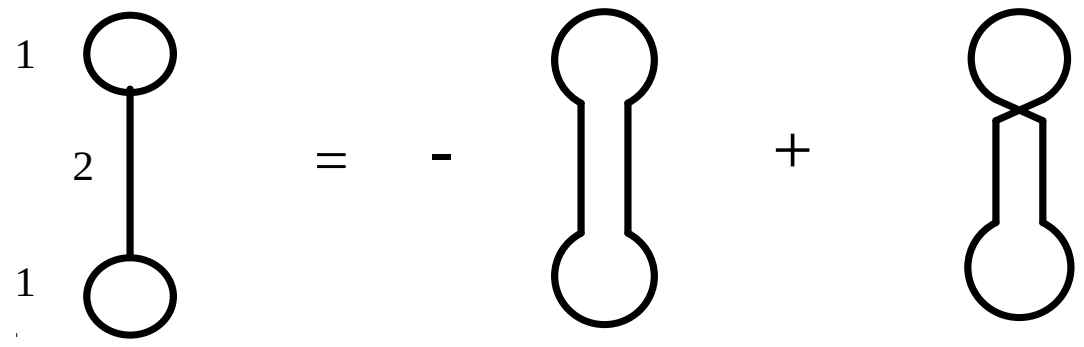

(24)

because for the first loop we have $c = 0, n = 1$ and for the second we have $c = 1, n = 1$; therefore the spin network represents the state

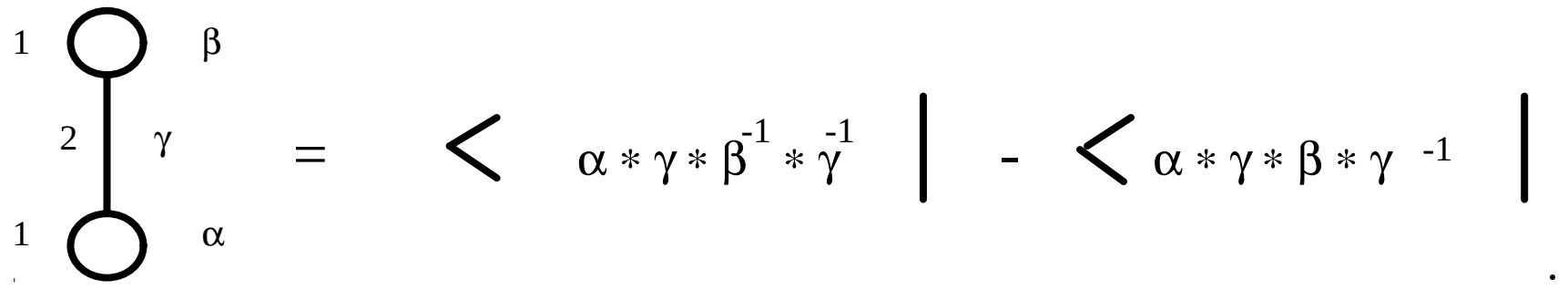

(25)

On the other hand, the spin network

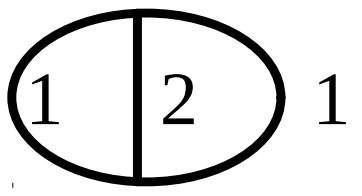

(26)

is expanded in loops as

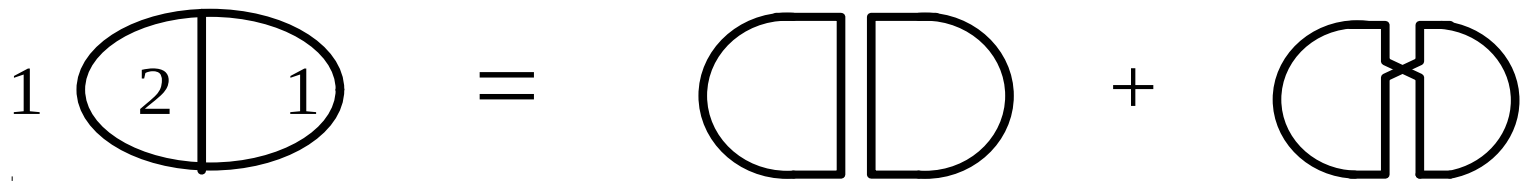

(27)

because we have $c = 0, n = 2$ for the first loop and $c = 1, n = 1$ for the second; therefore the spin network represents the state

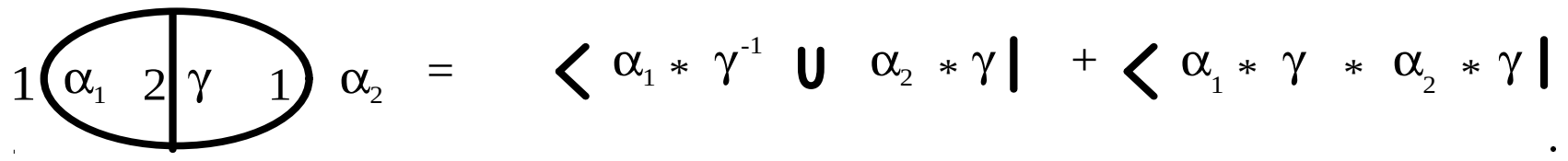

(28)

Notice the plus sign, contrary to the minus sign of the previous example.

The construction above can be easilly extended to loops with intesections of valence higher than 3. This is done by means of a simple generalization of the spin networks, obtained by considering non-trivalent graphs colored on the vertices al well as on the links. Or, equivalently, by trivalent spin networks in which sets of nodes are located in the same spacial point. This is worked out in detail in the Appendix.

Now, since the spin network states $\langle S|$ span the loop state space, it follows that any ket state $|\psi\rangle$ is uniquely determined by the values of the $\langle S|$ functionals on it. Namely, it is uniquely determined by the quantities

$$\psi(S) := \langle S|\psi\rangle. \tag{29}$$

Furthermore, since, as we shall prove later, the bra states $\langle S|$ are linearly independent, any assignement of quantities $\psi(S)$ corresponds to some ket $|\psi\rangle$. Therefore, quantum states in the loop representation can be represented by spin network functionals $\psi(S)$. By doing so, we can forget the difficulties due to the Mandelstam identities, which the loop states $\psi(\alpha)$ must satisfy.

In particular, we can consider spin networks characteristic states $\psi_T(S)$, defined by $\psi_T(S) = \delta_{T,S}$. We will later see that the Ashtekar-Lewandowski measure induces a scalar product in the loop representation under which the spin network states are orthonormal; then we can identify the characteristic states as the Hilbert duals of the spin network bra states. On the other hand, this identification depends on the scalar product, and thus in general one should not confuse the spin network characteristic states (kets) with the spin network states (bras).

It is easy to see that the calculations of the action of the Hamiltonian constraint $C$, presented in [2] imply immediately that if $\psi(S)$ vanishes on all spin networks $S$ which are not regular (formed by smooth and non self-intersecting loops), then $C\psi(S) = 0$. Notice that this follows from the combination of two results: the first is that $\langle S|C = 0$ for all regular $S$; the second is that $C\psi(S) = 0$ if $S$ is not regular; both these results are discussed in [2]. Thus, states $\psi(S)$ with support on regular spin networks solve the Hamiltonian constraint, and, at the same time, satisfy the Mandelstam identities. Indeed, they are precisely the extensions of the loop states with support on regular loops defined implicitely in [2] and discussed in detail in [26]. The spin network basis allows these solutions to be exhibited in a much more direct form.

The same conclusion may be reached using the form of the hamiltonian constraint described in [7], in which we consider the classically equivalent form of the constraint $\int_\Sigma f\sqrt{-C}$, where $f$ are smooth functions on $\Sigma$.

### 3.1 Diffeomorphism invariance and spin networks in knot space

One of the main reasons of interest of the loop representation of quantum gravity is the possibility of computing explicitely with diffeomorphism invariant states. These are given by the knot states. A knot $K$ is an equivalent class of loops under diffeomorphisms. We recall from [2] that a knots state, which we denote as $\psi_K$, or simply as $|K\rangle$ in Dirac notation, is a state of the quantum gravitational field with support on all the loops that are in the equivalence class $K$

$$\psi_K(\alpha) = \langle\alpha|K\rangle = \begin{cases} = 1 & \text{if } \alpha \in K, \\ = 0 & \text{otherwise.} \end{cases} \tag{30}$$

Clearly, the same idea works for the spin network states. Let us consider the equivalent classes of embedded oriented spin networks under diffeomorphisms. Such equivalence classes are entirely identified by the knotting properties of the embedded graph forming the spin network and by its coloring. We call these equivalence classes knotted spin networks, or s-knots for short, and indicate them with a lower case Latin letter as $s, t, r...$ . An s-knot $s$ can therefore be thought of as an abstract topological object independent of a particular imbedding in space, in the same fashion as knots. Then, for every knotted spin network $s$ we can define a quantum state $|s\rangle$ (a ket!) of the gravitational field by

$$\psi_s(S) = \langle S|s\rangle = \begin{cases} = 1 & \text{if } S \in s, \\ = 0 & \text{otherwise.} \end{cases} \tag{31}$$

Notice that in general a knot state $|K\rangle$ does not satisfy the Mandelstam relations. Diffeomorphism invariant loop functionals representing physical states should be constructed by suitable linear combination of the elementary knot states $|K\rangle$. This must be done by finding suitable extensions of the states from their values on regular knots to intersecting knots, using the Mandelstam identities, as described in [26]. However, these constructions are rather cumbersome, in spite of the fact that the rigorous results based on the use of diffeomorphism invariant measures on loop space [14] ensure us of the existence of such states. The spin network construction provides a way to circumvent this difficulty. Indeed, the s-knot state form a complete set of solutions of the diffeomorphism constraint; and the s-knot states corresponding to regular spin networks are solutions of all the constraints combined.

The space of the trivalent s-knots is numerable, for the same reason for which the set of the knots without intersections is numerable. However, we recall that diffeomorphism classes of graphs with intersections of order higher than five are continuous. To construct a separable basis for diffeomorphism invariant states including spin networks of all valences, a seperable basis must be selected for functions on each of these moduli spaces. As these spaces are finite dimensional, this can be accomplished. For a classification of the resulting moduli spaces of higher intersections, see [32].

This concludes the construction of the spin network states and of the s-knot states in the loop representation. To set the stage for the demonstration of their independence, we first define the spin network states in the connection representation.

## 4 The connection representation

We recall that in the connection representation one may consider a loop state $\psi_\alpha$, or $|\alpha\rangle$, defined by the trace[2] of the holonomy of the Ashtekar $SL(2,C)$ connection $A$ along $\alpha$,

$$\psi_\alpha(A) = \langle A|\alpha\rangle = T[A,\alpha] = Tr(U_\alpha) \tag{32}$$

Consider a spin network $S$. We can mimic the construction of the loop representation, and define the quantum state

$$\psi_S(A) = \langle A|S\rangle = \sum_m (-1)^{c(m)+n(m)} T[A,\gamma^S_m], \tag{33}$$

where, we recall, $n$ is the number of single loops and $c$ counts the terms of the symmetrization. Let us analyse this state in some detail. For every every link $l$ with color $p_l$, there are $p$ parallel propagators $U_l(A)$ along the link $l$, each one in the spin $1/2$ representation, that enter the definition of $\psi_S(A)$. Let us indicate tensors' indices explicitely; we introduce spinor indices $A, B, ...$ with value $0, 1$. The connection $A$ has components $A_A{}^B$, which form an $sl(2, C)$ matrix, and its parallel propagator along a link $l$ is a matrix $U_{l\,A}{}^B$ in the $SL(2, C)$ group. Since $SL(2, C)$ is the group of matrices with unit determinant, we have

$$\det U_l = \frac{1}{2}\epsilon_{AB}\,\epsilon^{CD}\,U_{l\,C}{}^A U_{l\,D}{}^B = 1, \tag{34}$$

where $\epsilon_{AB}$ is the totally antisimmetric two dimensional object defined by,

$$\epsilon_{01} = \epsilon^{01} = 1. \tag{35}$$

[2] Note that here we do not use the factor of 1/2 that has been conventional since the work of Ashtekar and Isham [13], but return to the original convention of ref. [2]. This choice is substantally more convenient for the present formalism, because otherwise we have to keep track of a factor of 1/2 for every trace, and these factors come into the relations between the spin networks and the loop states.

One can write $\psi_S(A)$ explicitely in terms of the parallel propagarors $U_{l\,A}{}^B$, the objects $\epsilon_{AB}$ and $\epsilon^{AB}$ and the Kroeneker delta $\delta_A^B$. Thus, any spin network state can be expressed by means of a certain tensor expression formed by $sl(2, C)$ tensors, $\epsilon$ and $\delta$ objects.

Penrose has described in [33] a graphical notation for tensor expressions of this kind. This notation is going to play a role in what follows, so we begin by recalling its main ingredients. We indicate two-indices tensors with thick lines, with the indices at the open ends of the line, respecting the distinction between upper indices, indicated by lines pointing up and lower indices, corresponding to lines pointing down. The sum over repeated indices is indicated by tying two lines together. More precisely, we indicate the matrix of the parallel propagator $U_{\alpha}{}_A^B$ of an (open or closed) curve $\alpha$ as a vertical bold line as in

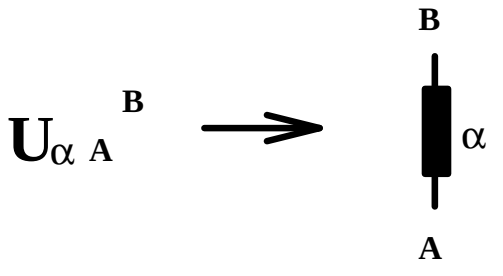

,

(36)

where the label $\alpha$ is understood unless needed for clarity; we indicate the antisymmetric tensors as in

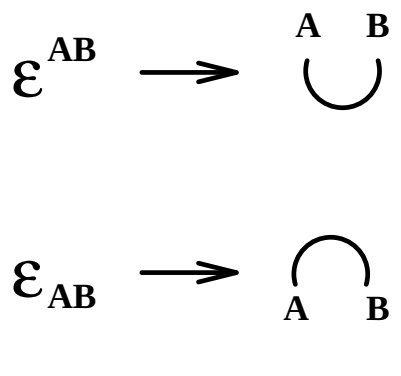

,

(37)

and the Kroneker delta as in

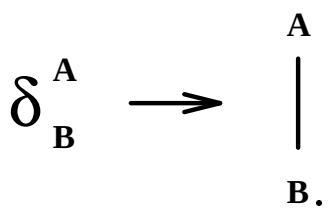

(38)

Finally, we indicate the sum over repeated indices by connecting the open ends of the lines where the indices are. We then have, for instance,

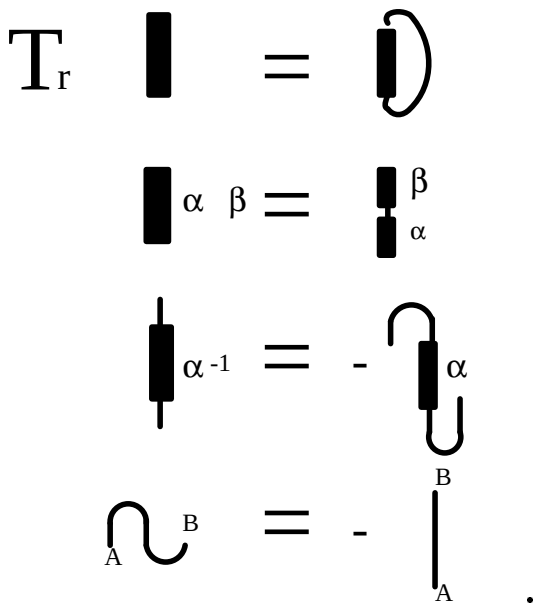

(39)

Also

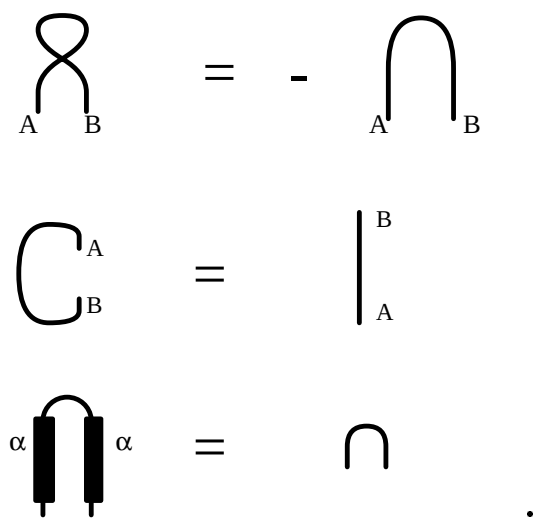

(40)

The most interesting relation is the identity

$$\delta_A^{\ B}\delta_C^{\ D} - \delta_A^{\ D}\delta_C^{\ B} = \epsilon^{BD}\epsilon_{AC}, \tag{41}$$

which becomes

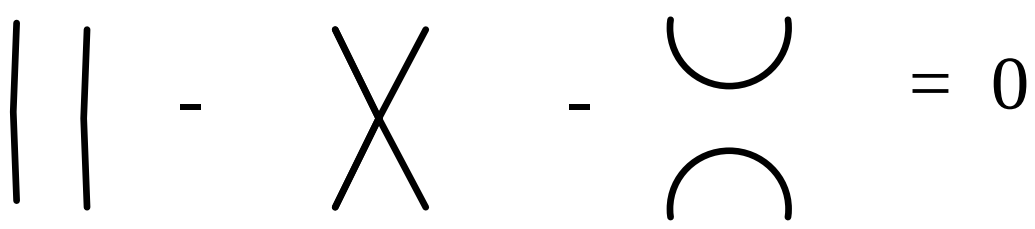

(42)

which is of course related to the loop representation spinor identity. Because of this last relation, in the Penrose diagram of any loop state we can use the

graphical relation

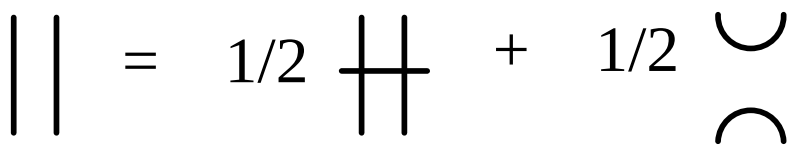

(43)

where the bar indicates simmetrization, on any (true) intersection, or overlapping loop.

Now, consider a generic (multiple) loop state in the connection representation; this is given as a product of terms, each of which is the trace of a product of matrices. We can represent these traces in terms of the corresponding graphical tensor diagram, which will result as a set of closed lines. We adopt the additional convention of drawing lines that form a rope as nearby parallel lines, and of reproducing the intersections of the original loops as intersections in the Penrose diagram (true intersections, representing intersections of the loop states, should be distinguished from accidental intersection forced by the planar nature of the Penrose diagram). In this way every multiple loop state is represented as a closed diagram. Let us denote this diagram as $G(\alpha)$ (for Graphical Tensor notation).

Notice that the diagram $G(\alpha)$ reproduces the topological features of the original loop $\alpha$; it can be naively thought as a simple two-dimensional drawing of the loop itself. But, the correspondence is not immediate, as is clear from the fact that the sign relations above imply that the same loop may correspond to either $G(\alpha)$ or to $-G(\alpha)$ depending on the way the loop is drawn.

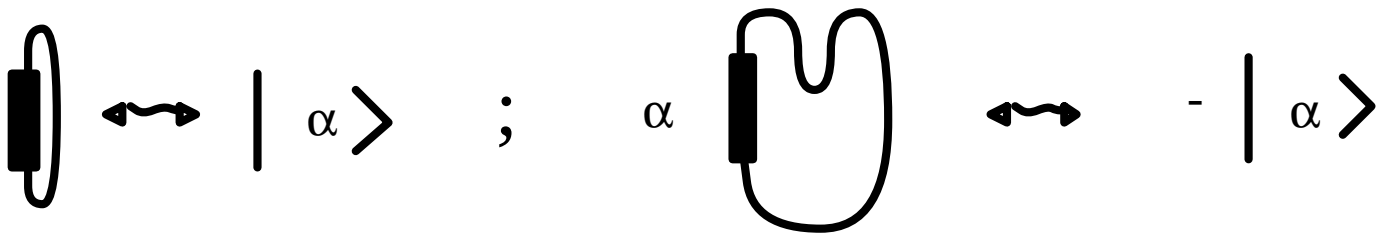

(44)

In order to distinguish between the drawing of the loop and the graphical tensor notation $G(\alpha)$ of the trace of the corresponding holonomy, from now on we denote the drawing of the loop as $D(\alpha)$. $G(\alpha)$ cannot be immediately identified with $D(\alpha)$. However, the relation between the graphical tensor diagram $G(\alpha)$ of the tensor $\psi_\alpha(A)$ and the planar representation $D(\alpha)$ of the loop $\alpha$ is not too difficult to work out. In fact, we have

$$G(\alpha) = (-1)^{m(\alpha)+c(\alpha)+n(\alpha)} D(\alpha) \tag{45}$$

where $m(\alpha)$ is the number of minima in the diagram $D(\alpha)$, $c(\alpha)$ is the number of crossings and $n(\alpha)$ is, as before, the number of single loop components of $\alpha$. This is an important formula, since it allows one to translate rigorously between graphical relations of the loop pictures and tensor relations of the corresponding holonomies. In a sense, this formula renders explicit an intuition that underlies the entire construction of the loop representation.

Let us work out the main consequence of this formula in the spin network context. The definition of the spin network states becomes, in tensor graphical notation

$$G(s) = \sum_m (-1)^{c(m)} (-1)^{n(m)+1} G(\gamma^s_m). \tag{46}$$

Therefore, expressing the right hand side in $D$ notation, we have

$$G(s) = \sum_m (-1)^{c(m)+1} (-1)^{n(m)+1} (-1)^{m(m)+c(m)+n(m)} D(\gamma^s_m) \tag{47}$$

or, simplifying the even exponents, noticing that the number of minima does not depend on the permutations, and absorbing an overall sign in the orientation

$$G(s) = \sum_m D(\gamma^s_m). \tag{48}$$

Thus we obtain the crucial conclusion that the tensor representing the spin network is obtained by writing one of the loop states, and consider all the permutations *with no sign factor*, namely by considering all *symmetrizations* of the lines along each rope. The resulting linear combination of graphical tensors gives directly the tensor representing the spin network state (up to an overall sign, that we can absorb in the orientation). Thus, we can conclude that the antisymmetrization that defines the spin network states is in fact a symmetrization of the SL(2,C) tensor indices. Let us now study what such a symmetrization implies

For every SL(2,C) tensor, we have from (34) the well known relation

$$\epsilon_{AB}\, \epsilon^{CD}\, U_C{}^A = (U^{-1})_D{}^B. \tag{49}$$

Consider a link $l$, and let $U_A{}^B$ be the paralell propagator of $A$ along $l$. Consider the product of two such propagators along the same link

$$U_{AC}{}^{BD} \equiv U_A{}^B U_C{}^D \tag{50}$$

This can be written as the sum of its symmetrized and antisymmetrized components

$$U_{AC}{}^{BD} = \frac{1}{2} U_{AC}{}^{(BD)} + \frac{1}{2} U_{AC}{}^{[BD]} \tag{51}$$

However, it is straightforward to show from the properties of two component spinors that

$$U_{AC}{}^{[BD]} = \epsilon_{AC}\epsilon^{BD}. \tag{52}$$

so that we have the identity

$$U_A{}^B U_C{}^D = \frac{1}{2} U_A{}^{(B} U_C{}^{D)} + \frac{1}{2}\epsilon_{AC}\epsilon^{BD} \tag{53}$$

If we write this in graphical tensor notation we have precisely equation (13). Following the same procedure, it is easy to show that a product of matrices $U_{B_1}^{A_1} U_{B_2}^{A_2} ... U_{B_n}^{A_n}$ can be decomposed in a sum of terms, each one formed by totally symmetrized terms $U_{B_1}^{(A_1} U_{B_2}^{A_2} ... U_{B_k}^{A_k)}$ times a product of epsilon matrices.

Of course what is going on here has a direct interpretation in terms of $SU(2)$ representation theory. Each matrix $U_A^B$ lives in the spin 1/2 representation of $SU(2)$; the product of $n$ of these matrices lives in the $n$-$th$ tensor power of the spin 1/2 representation, and this tensor product can be decomposed in the sum of irreducible representations. The irreducible representations are simply obtained by symmetrizing on the spin 1/2 indices. The reason we have reconstructed the details of the decomposition, is that this leads us to the precise relation between the tensorial expression of the connection representation states and the loop representation notation.

In fact, a fully antisymmetrized rope of degree $p$ is represented in matrix notation by a fully symmetrized tensor product of $p$ parallel propagators in the fundamental spin 1/2 representation. Therefore a rope of degree $p$ corresponds in the connection representation to a propagator in the spin $p/2$ representation. The result that every loop can be uniquely expanded in the spin network basis is equivalent to statements that the symmetrized products of the fundamental representation of $SL(2,C)$ gives all irreducible representations.

### 4.1 Nodes and $3j$ symbols: Explicit relations

The trivalent intersections between three ropes define an $SL(2,C)$ invariant product of three irreducible representations. Clearly the fact that there is a unique trivalent intersection in the loop representation is the reflection of the fact that there is a unique way of combining three irreducible representations to get the singlet representation, or, equivalently, that there is a unique decomposition of the tensor product of two irreducible representations. In this subsection we make the relation between the two formalisms explicit, for the sake of completeness.

Consider a triple intersection with adjacent lines colored $p, q$ and $r$. These correspond to the representations with angular momenta $l_p = p/2$, $l_q = q/2$ and $l_r = r/2$. The restriction on $p, q$ and $r$ that $p + q + r$ is even and none is larger than the sum of the other two corresponds to the basic tensor algebra relations of the algebra of the irreducible representations of $SL(2, C)$, namely the angular momentum addition rules. In fact, the two conditions are equivalent to the following familiar condition on $l_r$ once $l_p$ and $l_q$ are fixed:

$$l_r = |l_p - l_q|, |l_p - l_q| + 1, ...., (l_p + l_q) - 1, l_p + l_q. \tag{54}$$

Now, let $p, q$, and $r$ be fixed, and let us study the corresponding intersection in the connection representation. This is given by a summation over the symmetrized indices of the three products of parallel propagators. Let us raise all the indices of the propagators adjacent to the node. Let us denote by $U_B{}^A$ the spin 1/2 propagator along the link colored $p$, and by $V_B{}^A$ and $W_B{}^A$ the propagators along the links colored $q$ and $r$, where the propagators are oriented towards he intersection, so that the upstairs indices refer to the end on the intersection. Since the other index of each matrix is not going to play any role, we drop it, and write simply $U^A V^A$ and $W^A$. We must have at the intersection

$$U^{A_1} ... U^{A_p} V^{B_1} ... V^{B_q} W^{C_1} ... W^{C_r} K_{A_1 ... A_p, B_1 ... B_q, C_1 ... C_r} \tag{55}$$

where $K_{A_1 ... A_p, B_1 ... B_q, C_1 ... C_r}$ is an invariant tensor, symmetric in the first $p$ entries, the middle $q$ and the last $r$. Since the only invariant tensor is $\epsilon_{AB}$, $K_{A_1 ... A_p, B_1 ... B_q, C_1 ... C_r}$ must be a sum of products of $\epsilon_{AB}$'s. None of the $\epsilon_{AB}$'s can have both indices among the first $p$ indices of $K_{A_1 ... A_p, B_1 ... B_q, C_1 ... C_r}$, since $\epsilon_{AB}$ is antisymmetric and the first $p$ indices are symmetrised. Similarly for the middle $q$ and the last $r$. Thus we have $\epsilon_{AB}$'s with an $A$ index and a $B$ index (let's say we have $a$ of them) $\epsilon_{AB}$'s with a $B$ index and a $C$ index ($b$ of them) $\epsilon_{AB}$'s with a $C$ index and an $A$ index ($c$ of them). Clearly we must have $a + b = p, b + c = q$ and $c + a + r$. Thus $K_{A_1 ... A_p, B_1 ... B_q, C_1 ... C_r}$ may contain a term of the form

$$\epsilon_{A_{c+1} B_1} \cdots \epsilon_{A_p B_a} \quad \epsilon_{B_{a+1} C_1} \cdots \epsilon_{B_q C_b} \quad \epsilon_{C_{b+1} A_1} \cdots \epsilon_{B_r A_c} \tag{56}$$

We have to symmetrize this in each of the three set of indices. We obtain $p!q!r!$ terms, and it is not difficult to see that this sum is the only invariant tensor with the required properties. Thus

$$K_{A_1...A_p,B_1...B_q,C_1...C_r} = \sum \epsilon_{A_{c+1}B_1} \cdots \epsilon_{A_pB_a} \epsilon_{B_{a+1}C_1} \cdots \epsilon_{B_qC_b} \epsilon_{C_{b+1}A_1} \cdots \epsilon_{B_rA_c} \tag{57}$$

where the sum is over all the symmetrizations of the $A, B$ and $C$ indices. Now, notice that, if we read the graphical representation of the tensor as representing the loops, each of the terms in the sum corresponds precisely to the rooting of $a$ individual loops between the $p$ and the $q$ links, and so on. Thus, we obtain precisely the spin network vertex. On other hand, the relation between the matrix $K_{A_1...A_p,B_1...B_q,C_1...C_r}$ and the $3j$ symbols is also clear. For every representation with spin $l_p$, let us introduce the index $m_p$ that takes the $(2l_p + 1)$ values $m_p = -l_p, ..., l_p$. And in the basis $v_{mp}$ in the representation space related to the fully symmetrized tensor product of $2l_p$ spinors $\psi_{A_1}...\psi_{A_p}$, we write

$$v_{mp} = \psi_{(A_1} ... \psi_{A_p)} \sigma_{m_p}^{A_1...A_p} \tag{58}$$

Then we can write the vertex in this basis as

$$K_{m_pm_qm_r} = \sigma_{m_p}^{A_1...A_p} \sigma_{m_q}^{B_1...B_q} \sigma_{m_r}^{C_1...C_r} K_{A_1...A_p,B_1...B_q,C_1...C_r} \tag{59}$$

By uniqueness this must be proportional to the $3j$ symbols of SU(2):

$$K_{m_pm_qm_r} \sim \begin{pmatrix} l_p & l_q & l_r \\ m_p & m_q & m_r \end{pmatrix}. \tag{60}$$

# 5 Demonstration of the independence of the spin network basis

We are finally in the position to prove the independence of the spin network states $|s\rangle$. We will do this in the connection representation. The independence of these states is a linear property, and it should therefore be possible to prove it using only the linear structure of the space. However, it is much easier to construct a proof using an (arbitrary) inner product structure on the state space. Since linear independence is a linear property, once we have proven independence using a specific inner product, the result is independent from the inner product used.

Ashtekar and Lewandowski [14] have recently studied calculus on the space of connections, and have defined a measure $d\mu_{AL}(A)$ on (a suitable extension of) the space of connections, or, equivalently, a generalized measure on the space of connections [28]. Loop states, and their products are all measurable in this measure (in fact, the measure is defined using the technology of cylindrical measures, where the cylindrical functions are precisely the loop states.) Thus, the measure defines a quadratic form, or a scalar product, on the linear span of the loop states, which is finite as long as we consider only finite linear combination. For our purposes, the measure $d\mu_{AL}(A)$ is convenient for several reasons. First because it is diffeomorphism invariant. Second, because it is under good control, so calculations with it are easy. Here, we will use the original Ashtekar Lewandowski measure defined for $SU(2)$ connections. The extension to $SL(2,C)$ connections, is discussed in refs. [15]. In the present context, since loop states, as functionals on $SL(2,C)$ connections are holomorphic (functions of $A$ and not $\bar{A}$) they are determined by their restriction on the $SU(2)$ connections; thus, the $SU(2)$ measure that we employ defines an Hilbert space structure on these functionals, and this is all we need here. In any case, we refer the reader to reference refs. [15] for a much more accurate treatement of this point.

Let us thus consider functionals $f(A)$ of the connection of the form

$$f(A) = f(U_{\alpha_1}(A), ..., U_{\alpha_n}(A)), \tag{61}$$

where $f(g_1, ..., g_n)$ is a function on the $n-$th power of $SL(2,C)$, and thus, in particular on the $n-$th power of $SU(2)$. An example is provided by the loop states $\psi_\alpha(A)$. The $AL$ measure can be characterised as follows by

$$\int d\mu_{AL}(A) f(A) = \int f(g_1, ..., g_n)\ d_H(g_1, ..., g_n) \tag{62}$$

where $d_H$ is the Haar measure on the $n$-th power of $SU(2)$. The measure defines an inner product between loop states via

$$(\psi_\alpha, \psi_\beta) = \int d\mu_{AL}(A)\ \bar{\psi}_\alpha(A)\, \psi_\beta(A). \tag{63}$$

Now, what we want to prove is that the spin network states $\psi_s$ are linearly independent. Suppose that we can prove that they are all orthogonal with respect to this scalar product, namely

$$(\psi_s, \psi_s) \neq 0 \tag{64}$$

for every s, and

$$(\psi_s, \psi_{s'}) = 0 \qquad \text{for every } s' \neq s. \tag{65}$$

Then their linear independence follows, because if there were a linear combination of spin network states such that

$$\psi_s = \sum_m c_m \psi_{s_m} \tag{66}$$

we would have, taking the scalar product of the above equation with $\psi_s$ itself, a vanishing right hand side and a non-vanishing left hand side. Thus, to prove independence, we have to prove (64) and (65).

Let us consider a given spin network $s$. We have, using definitions

$$(\psi_s, \psi_s) = \int d\mu_{AL}(A) \bar{\psi}_s(A) \psi_s(A). \tag{67}$$

The spin network state is a sum (over permutations $m$) of products (over the single loops in the multiple loop $\gamma^s_m$ ) of traces of products (over the single links $l^j_{im}$ covered by the loop $\alpha_{im}$ ) of holonomies of $A_a$. Then

$$\psi_s = \sum_m (-1)^{n(m)+c(m)} \prod_i Tr \prod_j U_{l^j_{im}} \tag{68}$$

Therefore, using the definition of the Ashtekar-Lewandowski measure, we have

$$\begin{aligned}(\psi_s, \psi_{s'}) = \int dH(g(1)...g(L)) \sum_m (-1)^{n(m)+c(m)} \\ \times \prod_i Tr \prod_j g(l^j_{im}) \sum_{m'} (-1)^{n(m')+c(m')} \prod_{i'} Tr \prod_{j'} g(l^{j'}_{i'm'})\end{aligned} \tag{69}$$

where we have labelled as $1....L$ the links of the spin network. Now we have to use properties of the $SU(2)$ Haar measure. The main properties we need are

$$\begin{aligned}\int d_H(g) &= 1 \\ \int d_H(g)\ U(g)_A^{\ B} &= 0 \\ \int d_H(g)\ U(g)_A^{\ B} U(g)_C^{\ D} &= 12\ \epsilon_{AC}\epsilon^{BD}\end{aligned} \tag{70}$$

See reference [35] for a detailed discussion.

Let us analyze the effect of the integration graphically. Pick a link $l$, and consider the corresponding group integration $\int d_H(g(l))$. Let the colors of this link be $p$ and $p'$ for $s$ and $s'$. Then the integration over $g(l)$ is the integration over the product of $2p$ $U$'s. The result is the product of $\epsilon_{AB}$'s described above. Notice, however, that only the terms in which all the $\epsilon_{AB}$'s have one index coming from one of the spin networks and one index from the other can survive, because the others vanish due to the symmetry in the spin network indices and the antisymmetry in the result of the integration. Therefore, in each of the links involved in the integration there should be precisely the same number of segments in $s$ and $s'$. This is sufficient to see that any two spin network states corresponding to different spin networks are orthogonal.

Let us now take $s = s'$. Then, integrating the link $l$ gives a set of epsilons that connects the two copies of $s$ one with the other. We obtain thus terms in which, at every end of $l$ the two other adjacent links can simply be retraced back, plus terms which will vanish upon the next integration. Thus the two copies of $s$ get completely retraced back, leaving, at the end just products of integrations over the identity, each giving 1. Thus, we have shown that all spin network states are normalized. This completes the proof for trivalent states. The extension to higher valence intersections is simple; see also [28]. We may note that this result parallels the discussion of the independence of the spin network basis for hamiltonian lattice gauge theory, given, for example, in Furmanski and Kowala [19]. Given that the Ashtekar-Lewandowski measure is built from the projective limit of inner products for lattice gauge theories, for all analytic embeddings of lattices in $\Sigma$, it is not surprising that the result extends from lattice gauge theory to this case.

# 6 Relationship between the connection and the loop representation

In this section we review the relation between the loop representation $\mathcal{R}_l$ and the connection representation $\mathcal{R}_c$, which was introduced in [2]. This relation is simpler in the light of the spin network basis. To see this, we may recall from [2] that there exists a third relevant representation. This is the representation $\bar{\mathcal{R}}_c$ dual to the connection representation. The (ket) states in $\bar{\mathcal{R}}_c$ are, by definition, the bra-states of the connection representation, namely they are linear functionals on the space of functionals $\psi(A)$. Equivalently, we may think of these states as measures on the space of the connections. We denote the states in $\overline{\mathcal{R}}_c$ by $|d\mu\rangle$. The operators defining in $\mathcal{R}_c$ are immediately defined also in $\overline{\mathcal{R}}_c$, by their dual action.

In the absence of an inner product there is no canonical map between the state space of $\bar{\mathcal{R}}_c$ and the state space of $\mathcal{R}_c$. On the other hand, however, the representation $\bar{\mathcal{R}}_c$ is directly related to the loop representation $\mathcal{R}_l$. In fact, by double duality, any functional of the connection $\psi(A)$ defines a linear map on the state space of $\bar{\mathcal{R}}_c$, via

$$\langle\psi|d\mu\rangle = \overline{\langle d\mu|\psi\rangle} = \overline{\int d\mu(A)\ \psi(A)}. \qquad (71)$$

In particular, each loop state $|\psi_\alpha\rangle$, which is defined in the connection representation $\mathcal{R}_c$ by $\langle A|\psi_\alpha\rangle = T[A,\alpha]$, determines a dual state $\langle\psi_\alpha|$ in $\overline{\mathcal{R}}_c$, via

$$\langle\psi_\alpha|d\mu\rangle = \overline{\langle d\mu|\psi_\alpha\rangle} = \overline{\int d\mu(A)\ T[A,\alpha]} \qquad (72)$$

By definition of the loop representation [2], these bra states $\langle\psi_\alpha|$ in $\overline{\mathcal{R}}_c$ are to be identified precisely with the loop states $\langle\alpha|$:

$$\langle\alpha| = \langle\psi_\alpha|, \qquad (73)$$

In other words, the loop representation state $\psi(\alpha)$ is the state that is represented in the dual connection representation $\overline{\mathcal{R}}_c$ by the measure $d\mu$, where

$$\psi(\alpha) = \overline{\int d\mu(A) T[A,\alpha]}. \qquad (74)$$

This construction depends only on the linear structure of the quantum theory, namely it does not depend on a specific scalar product ( , ) which may or may not be defined on the state space. In the absence of a scalar product there is no canonical map between ket-space and bra-space, and therefore no canonical association of a dual state (a measure) $d\mu(A)$ to a given connection representation state $\psi(A)$; nor, equivalently, there is any canonical mapping between the connection representation $\mathcal{R}_c$ and loop representation $\overline{\mathcal{R}}_l$.

If a scalar product is given, then we can map connection representation states into loop representation states, because, given a connection representation functional $\psi(A)$, there is a unique dual state $d\mu_\psi(A)$ such that, for every $y'$,

$$\langle d\mu_\psi|\psi'\rangle = (\psi,\psi') \qquad (75)$$

And the loop state $\psi(\alpha)$ corresponding to the connection representation state $\psi(A)$ is then given by

$$\psi(\alpha) = \langle\psi_\alpha|d\mu_\psi\rangle. \qquad (76)$$

In particular, a scalar product in the connection representation can be assigned by fixing a measure $d\mu(A)$ on the space of connections, so that

$$(\psi, \psi') = \int d\mu(A)\overline{\psi(A)}\psi'(A). \tag{77}$$

Then, the induced map between the connection representation and the loop representation is the known expression for the loop transform

$$\psi(\alpha) = \int d\mu(A)\ \overline{T(A, \alpha)}\ \psi'(A). \tag{78}$$

To implement this relation explicitely we must use a measure $d\mu(A)$ which respects the invariances of the theory. Non-trivial gauge-invariant and diffeomorphism-invariant (generalised) measures on the space of connection have recently being constructed[14], and the Ashtekar-Lewandowski measure we used above is the simplest of these. The existence of this measure allows us to establish a definite linear map between the connection and the loop representation. Let us do so, and study its consequences. We will discuss later the extent to which we can take the resulting scalar product, and thus the resulting identification of the two representations, as the "physically correct" one.

For every ket state $\psi(A)$ in the connection representation, the Ashtekar-Lewandowski scalar product associates to it the bra state (or measure) $d\mu_\psi(A) = d\mu_{AL}(A)\psi(A)$. Let us consider a spin network state $\psi_s(A)$. The corresponding bra state state is $d\mu_s(A) = d\mu_{AL}(A)\psi_s(A)$. By the property of the spin network states under the Ashtekar-Lewandowski integration, we have then the remarkable result

$$(\psi_s, \psi_{s'}) = \langle d\mu_s | \psi_{s'}\rangle = \delta_{ss'}. \tag{79}$$

The loop representation state bra state $\langle s|$ corresponding to $\psi_s$ is defined solely in terms of the linear properties of the representation. The loop representation ket state $|s\rangle$, on the other hand, is defined by

$$\psi_s(\alpha) = \overline{\int d\mu_s(A) T[A, \alpha]}. \tag{80}$$

and it follows immediately that it is the adjoint of $\langle s|$. Thus, the Ashtekar Lewandowski inner product becomes, in the loop representation:

$$\langle s|s'\rangle = \delta_{ss'}, \tag{81}$$

a remarkable result indeed. In other words the spin network basis is orthogonal with respect to this inner product. It is important to notice that these results do not hold for the loop states themselves, for which inner products of the form $\langle\alpha|\alpha'\rangle = \delta_{\alpha\alpha'}$ are inconsistent with the Mandelstam relations.

Finally, let us note that the condition on the theory's physical inner product is that it makes the real observables hermitian. While little is known about the solution of this condition, because of the absence of explicit operator expressions for physical observables, we may say more at the kinematical and spatially diffeomorphism invariant level. At the kinematical level, we may note that the volume and area operators are enough to distinguish all spin networks from each other. This means that they must be orthogonal to each other in any kinematical inner product. If we impose the additional conditions that the loop operators $T[\alpha]$ be hermitian, which means that we are describing Euclidean quantum gravityor a formulation such as that in [34] in which the connection is real, then it is straightforward to show that, at this kinematical level, the spin network states have unit norms. An important open problem is whether there is a different choice of norms for the spin network basis that realizes the Minkowskian reality conditions.

At the diffeomorphism invariant level, we may note that we have at least one explicit observable, which is the volume of the spatial slice $\Sigma$. If this is to be hermitian then we know that the spin network states corresponding to different volumes must be orthogonal to each other. This is related to the fact that both the area and volume [10] are indeed symmetric with respect to the Ashtekar-Lewandowski inner product. We may note that the diffeomorphism invariant inner product is physically meaningful in the treatment of evolution described in [7, 11]

# 7 Penrose diagram notation

Our final task in this paper is to exploit the results we have described to introduce a notation for the loop states, which simplifies the graphical calculus in the loop representation. We recall that we indicate as $D(\alpha)$ the pictorial representation of the loop $\alpha$. We are now going to define a notation for the loop states, which we denote as the Penrose notation, or $P$ notation. (This was also called the binor formalism by Penrose.) This is done as follows. In $P$ notation a loop state $|\alpha\rangle$ is also represented by a certain pictorial representation of the loop itself, which, to distinguish it, we will call $P(\alpha)$. However, the representation takes into account the sign factor that we discussed in previous sections. Thus, we choose the convention that the diagram of a loop, $P(\alpha)$ in $P$ notation is related to the loop state $|\alpha\rangle$ by $(-1)^{n(\alpha)+1}|\alpha\rangle$, where $n(\alpha)$, we recall, indicates the number of single loops, or components, of the multiple loop $\alpha$. In other words, the $P$ notation $P(\alpha)$ of a loop state $|\alpha\rangle$ is defined by

$$P(\alpha) = (-1)^{n(\alpha)+1} D(\alpha). \tag{82}$$

The important aspect of the $P$ notation is that with these conventions the the spinor identy is now local. In fact it now reads as

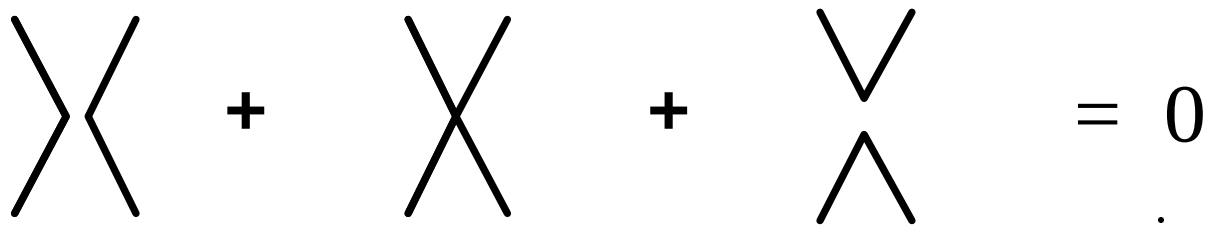

(83)

Therefore equation (13) holds rigorously whithin this notation. Thus can solve the spinor identity (on trivalent states) by restricting to the states in which every rope is (in $P$ notation) fully antisymmetrized. It is important to note that the $P$ notation is completely topological, in that a diagram corresponds to the same loop state no matter how it is oriented ot drawn. This is a great advantage in calculations.

For completeness we mention a variant of the $P$ notation, has been used in some published work [10]. The variant, which we may denote Symmetric Penrose notation, corresponds to what Penrose called the spinor calculus, as opposed to binor calculus. In this notation, which we will refer to as $SP$ notation, the diagram that corresponds to a loop $\alpha$ will be labled $S(\alpha)$. It is defined by,

$$S(\alpha) = (-1)^{n(\alpha)+c(\alpha)+m(\alpha)+1} D(\alpha). \tag{84}$$

We recall that $c(\alpha)$ and $m(\alpha)$ are the number of crossings and the number of minima in $D(\alpha)$. Notice that the $SP$ notation is not "topological", in the sense that the way the loop a is drawn matters for the determination of the sign: adding a minumun and a maximum is equiavelent to change the sign of the state. Notice that the permutations of the loops along a rope change the number of crossings therefore the antisimmetrization in the $P$ notation corresponds to a symmetrization in $SP$ notation (hence the name). While it is more cumbersome for calculation, the significance of the $SP$ notation is that it has an immediate interpretation in terms of Penrose graphical tensor calculus, which we defined earlier in the connection representation. Indeed,

we have immediately that

$$SP(\alpha) = G(\alpha). \tag{85}$$

Finally, we mention the fact that the $P$ notationcan be obtained from the graphical tensor notation by adding the immaginary unit to each $\epsilon$, and adding a minus one for every crossing.

### 7.1 Loop operators in Penrose notation

A most valuable aspect of the Penrose diagram notation we have introduced is the simplification it allows in the calculus with the loop operators. In this section we describe the action of the loop operators on the loop states expressed in Penrose notation.

We recall [2] that in terms of the standard loop notation the action of the loop operators $T^1$ and $T^2$ is given by

$$\langle\beta|T^a[\alpha](s) = \Delta^a[\beta,\alpha(s)](\langle\alpha\cdot\beta| - \langle\alpha\cdot\beta^{-1}|) \tag{86}$$

$$\langle\beta|T^{ab}[\alpha](s,t) = \Delta^a[\beta,\alpha(s)]\Delta^b[\beta,\alpha(t)]\sum_i(-1)^{r(i)}\langle(\alpha\cdot_{st}\beta)_i| \tag{87}$$

The distributional factor $\Delta^a[\beta,\alpha(s)]$ (which doesn't play any role in the diagramatics) is defined in [2]. The geometrical part of the action of these operators can be coded in the "grasping" action shown in

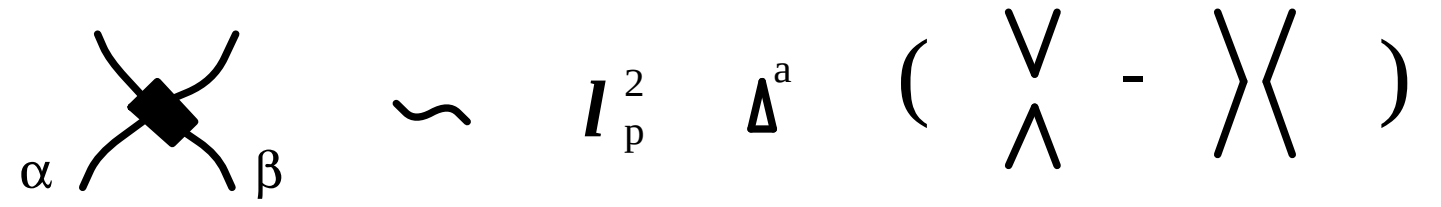

(88)

However, as is well known by anybody who attempted to perform complex computations with these operators, the local graphical action expressed in eq. (88) does not suffice to compute the correct linear combination appearing in the r.h.s of eq. (87). The difficulty is given by the signs in front of the various terms. These signs are dictated by the global rooting properties of the loop that are being grasped. In particular the sign is determined in (87) by $r(i)$, which is defined [2] as the number of segments that have to be reversed in order to obtain a consistent orientation of the loop after the rerooting. While complete, this way of determining the sign is cumbersome, and in computing the action of operators as the area, the Hamiltonian, or the volume, the determination of the signs is the hardest part of the calculation. This difficulty disappears using the Penrose diagram notation.

Let us begin by considering the action of $T^1$ on a loop state.

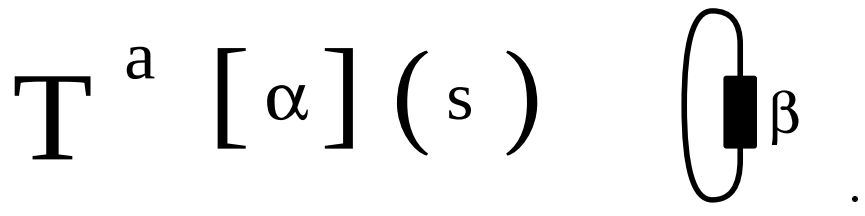

(89)

To compute this we express the r.h.s of in Penrose notation. The result is

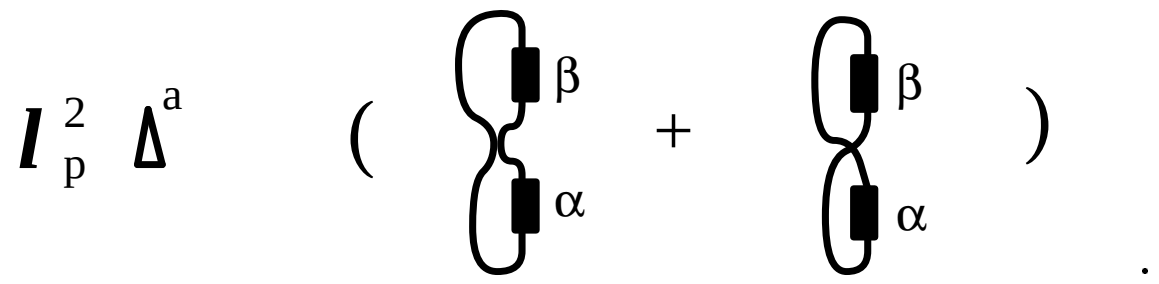

(90)

Notice the plus sign, due to the change of sign

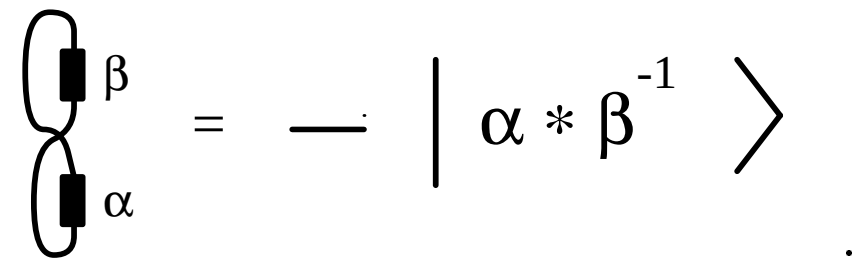

(91)

This suggests that we indicate the operator $T^a[\alpha](s)$ by

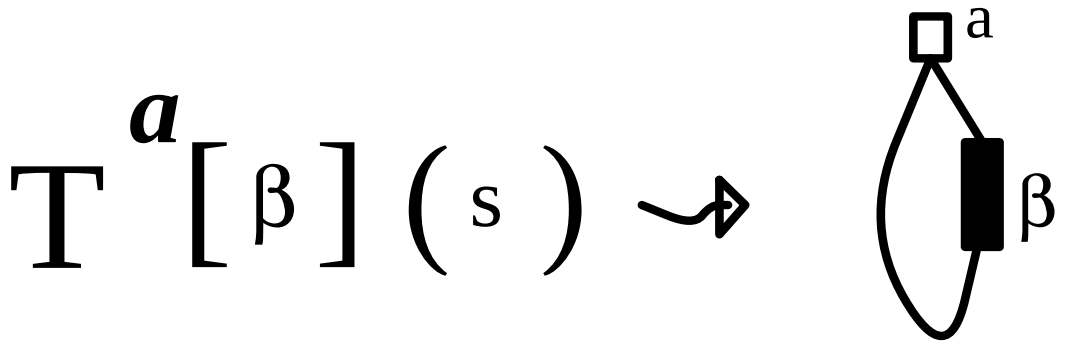

(92)

To notate the action of the operator we may then introduce the following fundamental "grasping" rule

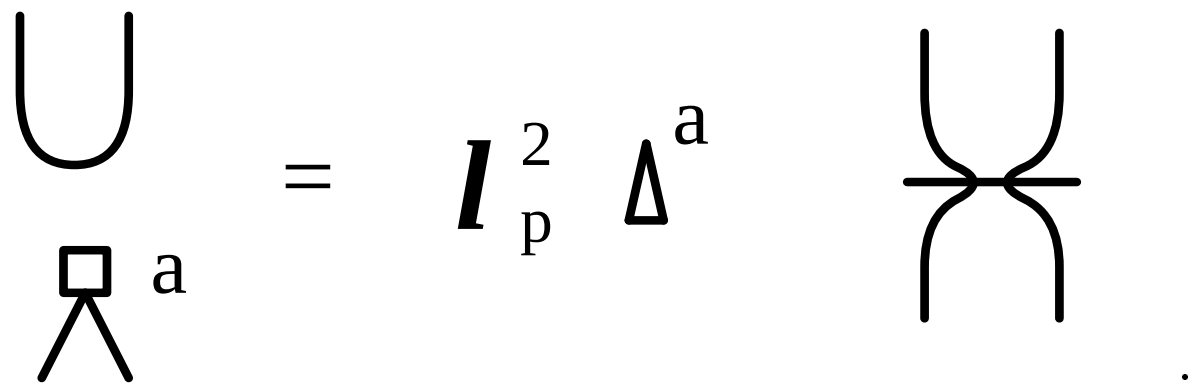

(93)

By using this rule we have immediately the correct action, represented as

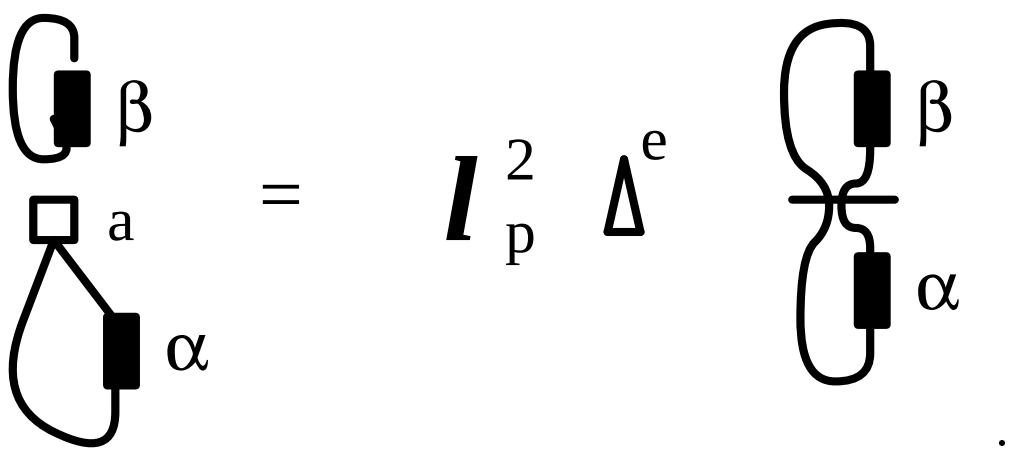

(94)

The good news are that this generalizes immediately to the higher order loop operators. For example, let us represent the $T^2$ loop operator as

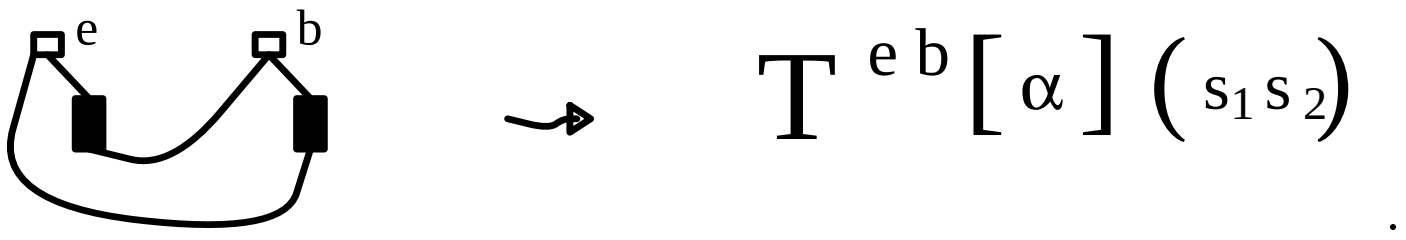

(95)

Then we can use the fundamental grasping rule above to compute the action of $T^2$ on a generic state. We obtain

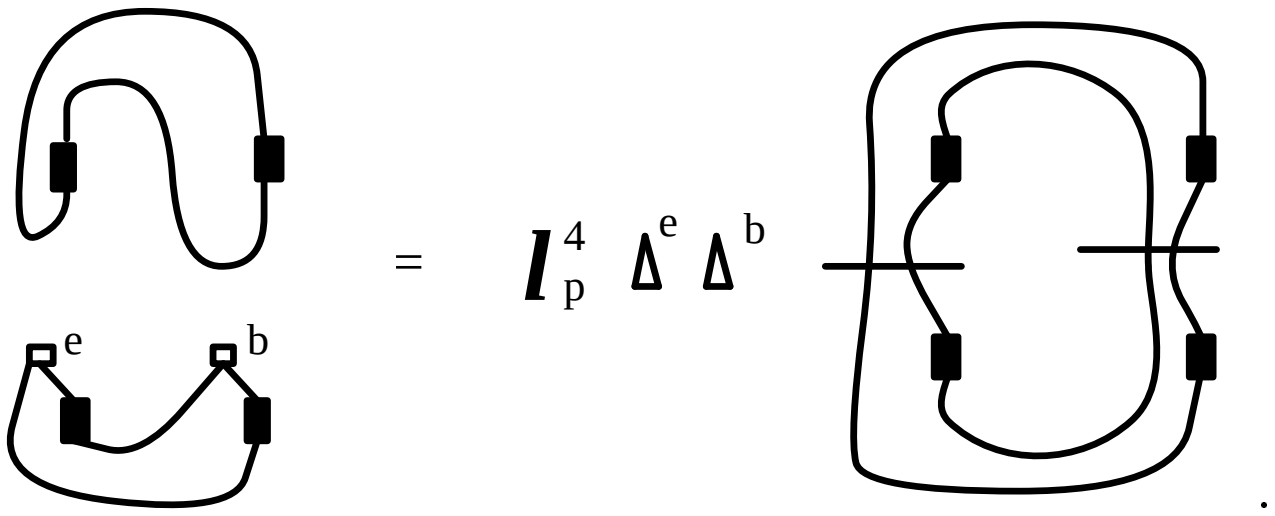

(96)

By expanding the diagrams, and taking into account the sign rules, it is straightforward to show that the r.h.s of this diagram represents the correct linear combination of loop states, corresponding to the r.h.s of equation (87). This result generalizes to higher order $T^n$ operators. Thus, by using the Penrose notation the grasping rule of eq. (93) encodes automatically the pattern of the signs in the action of the $T^n$ operators.

These simplifications extend to all the higher $T^n$ operators. For example, in [10] we showed how this this notation simplifies the computation of the action of the volume operator which is defined in terms of a $T^3$ operator. This made much simpler the work of solving the corresponding spectral problem, leading to the computation of the eigenstates and eigenvalues of the operator that corresponds to the volume of an arbitrary region of space.

## 8 Conclusion

We have defined a basis of independent states in the loop and in the connection representations of quantum gravity which solves the Mandelstam identities. This basis is labelled by a generalization of Penrose spin networks. It is orthonormal in the scalar product defined by the Asktekar-Lewandowski measure, and provides a simple relation between the connection and the loop representation. We have introduced a notation for the loop states of quantum gravity based on Penrose's graphical tensor notation. In this notation, the action of the loop operators becomes local, and can be expressed in terms of the simple graphical rule given in equation (93).

An intriguing suggestion on the possibility of modifying the framework we have presented in this paper follows from the following observation. Because of the short-scale discretness of the geometry [10], the only remaining divergences in nonperturbative quantum gravity must be infrared divergences, analogous to the spikes, or the uncontrolled proliferation of "baby universes" seen in nonperturbative numerical calculations employing dynamical triangulations [36] in both two and four dimensions. In the present context, a source of such divergences may be the sum over the colorings of the spin networks, which label the representations of $SU(2)$. This suggests that a natural invariant regularization of the theory could be provided by replacing $SU(2)$ with the quantum group $SL(2)_q$. Such a strategy has been successfully implemented in 3 dimensions by Turaev and Viro [21], and there have been attempts to extend it to four dimensional diffeomorphism invariant theories [23]. The use of q-deformed spin networks in the loop representation of quantum gravity is presently under investigation [37]. Furthermore, spin networks may make it possible to define quantum gravity on manifolds with a finite boundary [38], and to use the methods of topological field theory to describe the structure of the physical quantum gravity state space in the presence of boundaries. In this context, the level $q$ of $SL(2)_q$ turns out to be related to the inverse of the cosmological constant [38]. These investigations reinforce the conjecture that the $q$ deformation could play the role of infrared regulator. The possible relevance of $q$ deformations of the gauge group $SU(2)_L$ in quantum gravity is also suggested by the important role quantum groups play in knot theory [39], as well as by the possibility of quantum-gravity induced, quantum-group deformations of the space-symmetries [40].

Finally, we remark that the existence of a spin network basis for the space of diffeomorphism invariant states of the quantum gravitational field, as well as the important role they seem to play in both practical calculations [10, 11, 12, 38] and mathematical developments [28, 15, 27] may be seen as vindicating the picture of a discrete, combinatorial description of spacetime geometry, as well as the reasoning that led Roger Penrose to their original construction [16].

## Acknowledgements

We thank Roger Penrose for introducing us to the details of the spin network calculus and John Baez and Louis Kauffman for explanations and various suggestions about the use of spin networks. We also thank Roumen Borissov for many comments and suggestions, Abhay Ashtekar, Bernd Bruegmann, Louis Crane, Junichi Iwasaki, Renata Loll, Seth Major, Jorge Pullin and other members of the Center for Gravitational Physics and Geometry for discussions as well as for comments on an earlier draft of this paper. We are grateful to Harlow Igims for help with the figures. Finally, ls would like to thank Prof. Chris Isham and the Newton institute for support for a visit during which this work was begun, and Prof. Mauro Carfora and SISSA for support and hospitality, where it was continued. This work was partially supported by the NSF under grants PHY539634, PHY9016733, PHY9311465 and INT8815209 and by research funds of Penn State University.

## Appendix

In this Appendix we extend the definition of spin network states to intersections of any order. The complication introduced by higher order intersections is the fact that the order of the ropes entering the intersection does not determine the routing uniquely. For instance, in the simplest possible fourth order intersection, with all four ropes of order 1, we have three possible rootings

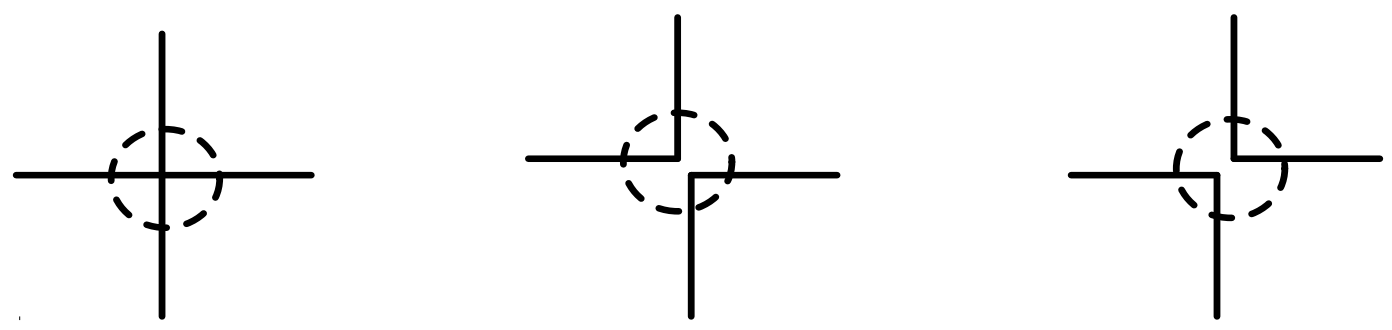

(97)

out of which two are independent, due to the spinor identity. Nevertheless, with a small amount of additional technical machinery, it is possible to extend the spin network basis to include arbitrary intersections. This is because given the order $n$ of an intersection $i$, and given the coloring $p_1, ..., p_n$ of the $n$ ropes adjacent to $i$, there is only a finite number of ways of rootings the loops through the intersection, and therefore a (smaller) finite number $k(p_1, ..., p_n)$ of *independent* rootings. For completeness, we put $k(p_1, ..., p_n) = 0$ if a consistent rooting through the intersection does not exist for $n$ ropes of orders $p_1, ..., p_n$; this is for instance the case if $\sum_j p_j$ is odd. In the particular case of trivalent intersections ($n = 3$) we have $k(p_1, ..., p_3) = 1$ if the sum of three colors $p_j$ is even and none of the three is larger than the sum of the other two, and $k(p_1, ..., p_3) = 0$ otherwise.

In order to extend the definition of spin network states to non-trivalent loops, it is sufficient to choose a unique way of labeling the $k(p_1, ..., p_n)$ independent rootings through an intersection $i$, by means of an integer $v_i = 1, ..., k(p_1, ..., p_n)$. Once this is done, we define a generalized spin network $s$ as an oriented imbedded graph $\Gamma$, with positive integers, or colors, $p_l$ and $v_i$ assigned to each link $l$ and to each of node $i$; satisfying the relations $v_i \leq k(p_1, ..., p_n)$, $p_1, ..., p_n$ being the colors of the links adjacent to the node $i$. The construction of the corresponding spin network quantum states $\langle s|$ is then as before.

The task of labeling independent spin networks can be achieved as follows. For every $n$, we choose a unique trivalent graph $\Gamma^{(n)}$ with n free ends, and no closed loop; for instance, we may choose

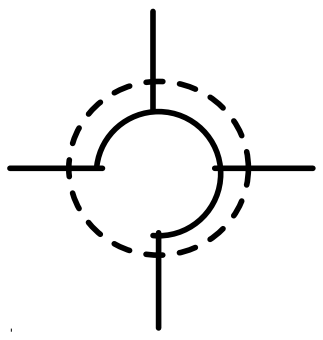

(98)

Such a graph will have $n$ links adjacent to the free ends, and $(n-3)$ internal links, which we denote as "virtual" links. For every $n$, and every set of colors $p_1, ..., p_n$ we consider the possible colorings $q_1, ..., q_{(n-3)}$ of the virtual links of $\Gamma^{(n)}$ which are compatible with the colorings $p_1, ..., p_n$ of its external links (under the spin networks vertex conditions). We obtain in this way a family of colored trivalent spin networks $\Gamma^{(n)}_1, ..., \Gamma^{(n)}_{k(p_1,...,p_n)}$ with $n$ external links colored $p_1, ..., p_n$. It is not difficult to see that these are linear combinations of rootings through the intersection which exhaust all possibilities, up to the spinor identities and which are indipendent from each other. We label these intersection with the integer $v_1 = 1, ..., k(p_1, ..., p_n)$. If there is no way of matching the coloring we put $k(p_1, ..., p_n) = 0$.

Let us work out an example of fourth order intersection.

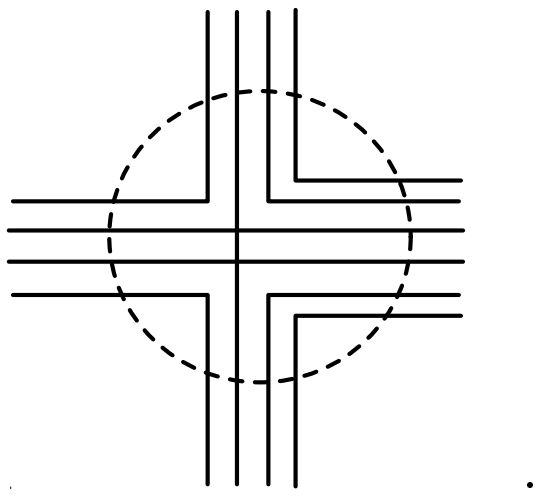

(99)

We arbitrarily pair the four ropes; for instance, let us pair the North and West ropes and the South and East ones, and "expand" the intersection by introducing an additional ("virtual") rope between the joins of the paired ropes: We obtain

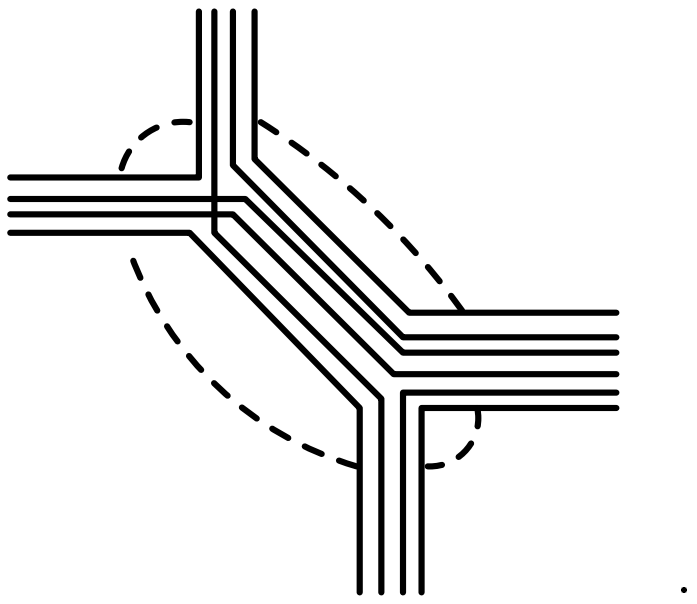

(100)

In this way, the fourth order intersection is "expanded" into two trivalent intersections. Notice that in Figure 12 the external ropes are symmetrized, while the internal one is not. By using the spinor relations, we can then replace the diagram with a linear combination of diagrams in which the internal rope too is symmetrized. Thus, we can represent the intersection

as

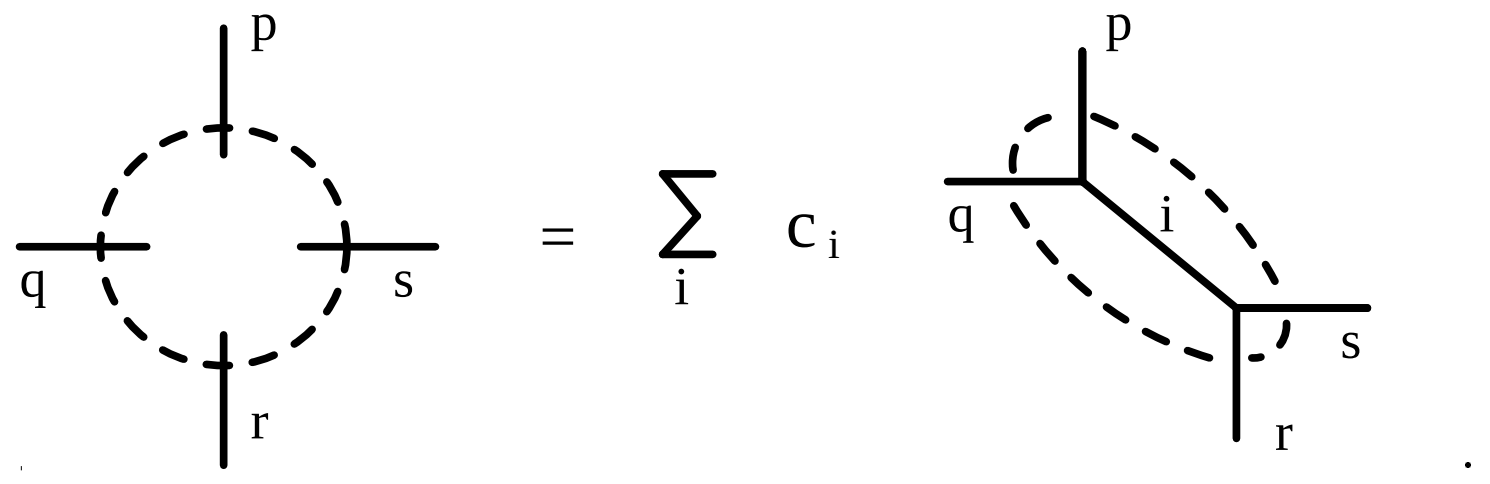

(101)

where we have used the spin network notation. The coefficients $c_i$ depend on (and can be computed from) the original rootings in the fourth order intersection. The index $i$ ranges from $max(|p-q|,|r-s|)$ to $min(p+q,r+s)$. Finally, once the pairing is chosen, it is clear that the decomposition of eq. (13) is always possible and unique, and it reduces the spinor identities completely. Thus, we have

$$k(p,q,r,s) = max(|p-q|,|r-s|) - min(p+q,r+s) \tag{102}$$

independent fourth order intersections between ropes of orders $(p,q,r,s)$, and we have a simple way of ordering them in terms of the color of the internal rope.